\documentclass[11pt,a4paper]{article}
\usepackage{geometry}
\usepackage{graphicx}
\usepackage{bm,array}
\usepackage{comment}
\usepackage{caption}
\usepackage{subcaption}
 \usepackage{makecell}
 \usepackage{multirow}
 \usepackage{float}

\usepackage[normalem]{ulem}

\usepackage[pdfpagemode={UseOutlines},bookmarks=true,bookmarksopen=true,
bookmarksopenlevel=0,bookmarksnumbered=true,hypertexnames=false,
colorlinks,linkcolor={blue},citecolor={blue},urlcolor={red},
pdfstartview={FitV},unicode,breaklinks=true]{hyperref}
\hypersetup{urlcolor=blue, colorlinks=true}

\usepackage{setspace}
\usepackage{vmargin}
\setmarginsrb{ 1.0in}  
{ 0.6in}  
{ 1.0in}  
{ 0.8in}  
{  20pt}  
{0.25in}  
{9pt}  
{ 0.3in}  

\usepackage{authblk}

\usepackage{amsmath}
\usepackage{amsfonts}
\usepackage{amssymb}
\usepackage[round, sort,comma,authoryear]{natbib}

\newtheorem{theorem}{Theorem}

\newtheorem{corollary}{Corollary}

\newtheorem{lemma}{Lemma}

\newtheorem{proposition}{Proposition}

\newenvironment{proof}[1][Proof]{\noindent\textbf{#1.} }{\ \rule{0.5em}{0.5em}}

\usepackage{multirow}
\usepackage{hhline}
\usepackage{caption}
\usepackage{subcaption}
\usepackage{footnote}
\usepackage[ruled,vlined]{algorithm2e}
\usepackage{algorithmic}
\usepackage{soul}

\newcommand{\transpose}{{\mbox{\tiny T}}}

\newcommand{\cA}{{\mathcal{A}}}

\newcommand{\cG}{{\mathcal{G}}}

\newcommand{\cD}{{\mathcal{D}}}

\newcommand{\cL}{{\mathcal{L}}}

\newcommand{\cN}{{\mathcal{N}}}

\newcommand{\cT}{{\mathcal{T}}}

\newcommand{\cO}{{\mathcal{O}}}

\newcommand{\bI}{{\textbf{I}}}

\newcommand{\bM}{{\textbf{M}}}

\newcommand{\bx}{\textbf{x}}

\newcommand{\ba}{\textbf{a}}

\newcommand{\bz}{\textbf{z}}
\newcommand{\bb}{\textbf{b}}

\newcommand{\bbt}{\pmb{\beta}}

\newcommand{\bbE}{\mathbb{E}}

\newcommand{\BiC}{{\sf \tiny BiC}}
\newcommand{\MuC}{{\sf \tiny MuC}}

\usepackage{mathtools}

\usepackage{mathtools}

\newif\ifnotes\notestrue
\def\htien#1{}

\begin{document}






\newcolumntype{C}{>{\centering\arraybackslash}p{4em}}

\title{\textbf{Network-based Representations and Dynamic Discrete Choice Models for Multiple Discrete Choice Analysis}}
\author[1]{Hung Tran}
\author[2,*]{Tien Mai}
\affil[1]{\it\small
School of Computing and Information Systems, Singapore Management University, 80 Stamford Rd, Singapore,
{hhtran@smu.edu.sg}}
\affil[2]{\it\small
School of Computing and Information Systems, Singapore Management University, 80 Stamford Rd, Singapore, 
{atmai@smu.edu.sg}}
\affil[*]{\it\small Corresponding author}

\maketitle

\begin{abstract}
In many choice modeling applications, people demand is frequently characterized as multiple discrete, which means that people choose multiple items simultaneously. The analysis and prediction of people behavior in multiple discrete choice situations pose several challenges. In this paper,  to address this, we propose a random utility maximization (RUM) based model that considers each subset of choice alternatives as a \textit{composite} alternative, where individuals choose a subset according to the RUM framework. While this approach offers a natural and intuitive modeling approach for multiple-choice analysis, the large number of subsets of choices in the formulation makes its estimation and application intractable. To overcome this challenge, we introduce directed acyclic graph (DAG) based representations of choices where each node of the DAG is associated with an elemental alternative and additional information such that the number of selected elemental alternatives. Our innovation is to show that the multi-choice model is equivalent to a \textit{recursive route choice model} on the DAG, leading to the development of new efficient estimation algorithms based on dynamic programming.
In addition, the DAG representations enable us to bring some advanced route choice models to capture the correlation between subset choice alternatives. 
Numerical experiments based on synthetic and real datasets show many advantages of our modeling approach and the proposed estimation algorithms.
\end{abstract}







%


\section{Introduction}\label{sec:intro}
The demand of consumers is often characterized as multiple discrete, where they choose several items simultaneously. For example, a household may choose a mix of different vehicle types, an individual may participate in multiple daily activities, a tourist may purchase multiple air tickets for their family, or a consumer may purchase multiple items from a store or borrow several books from a library. In the choice modeling literature, researchers have mentioned various ways to handle multiple discrete choice behaviors. The most natural and intuitive way would be to use the traditional random ultility maximization (RUM) \citep{McFa81} single discrete choice models by identifying all combinations of elemental alternatives and treating each combination as a composite alternative. However, this approach is known to be intractable to use, as the number of composite alternatives grows exponentially as the number of elemental alternatives grows. 
Other modeling approaches include the multiple discrete-continuous models (MDC) \citep{bhat2008multiple,bhat2015allowing}. Such models have been proposed to address choice situations that involve a subset of elemental alternatives along with continuous intensities of consumption for the chosen elemental alternatives. While such models use continuous values to present the number of choices, they can get rid of the combinatorial structure of the choice model and the issue of having exponentially many choice alternatives. However, they cannot handle multiple discrete choice situations. Extensions have been proposed to handle the discrete nature of people's choice behavior \citep{lee2022sequential,bhat2022closed}, but these models are still expensive to estimate and perform prediction. In fact,
existing approaches 
 are typically based on constrained utility maximization, which assumes that a consumer finds an exact optimal solution that maximizes their overall utility while satisfying some complex constraints. This can be a strong assumption since constrained utility maximization is not an easy problem to the choice-makers. Furthermore, it also conflicts with the importance  of bounded rationality in
decision-making, stated in several prior cognitive psychology and behavioral
economics studies \citep{sent2018rationality,campitelli2010herbert}.

In this paper, we get back to the RUM-based approach, as it would be one of the most natural and intuitive ways to model multiple discrete choice behaviors. Specifically, we focus on addressing the computational challenge that arises when there is an exponential number of subset choice alternatives to be considered. 
To overcome this challenge, we draw inspiration from the observation that the multiple discrete choice problem shares some synergies with the route choice modeling context, where the number of paths going from an origin to a destination in a network is exponentially large and it is almost infeasible to enumerate all the paths and use them. We, therefore, bring some advanced route choice models and algorithms to address the challenge.  Our approach involves the proposition of Directed Acyclic Graphs (DAG) to represent the multiple choice process, where each node can represent an elemental alternative with some additional information, such as the number of selected alternatives. We then show that there is an equivalence between choosing multiple choices and walking through the DAG. This approach allows us to leverage dynamic programming techniques to efficiently estimate the multiple discrete choice model. Further, we also demonstrate that some advanced route choice models, such as the nested recursive logit \citep{MaiFosFre15}, can be used to capture complex correlation structures between composite choice alternatives. We summarize our main contributions in the following.

\noindent
\textbf{Contributions:} We make the following contributions.

We introduce a RUM-based model for multiple discrete choices where each individual is assumed to consider a set of composite alternatives and make a choice according to a standard RUM model, such as the multinomial logit (MNL) model \citep{Trai03}. An important aspect of our model is that the choice-maker is subject to a constraint on the range of the size of the composite alternative. Such a constraint is reasonable and simple enough to be considered in the context.

To handle the estimation problem, which involves an exponential number of composite alternatives, we propose two DAG-based representations for the choices. Each node of the DAG represents an elemental alternative and can include some additional information, such as the number of selected elemental choices. By employing the recursive logit (RL) route choice model \citep{FosgFrejKarl13}, we show that the probability of selecting a composite alternative is equal to the probability of selecting a path in the DAG. This allows us to use results from the route choice modeling literature to show that the choice probabilities of composite alternatives can be computed by solving systems of linear equations, which can be done in poly-time. Consequently, the computation of the log-likelihood can also be done in poly-time. It is noted that the DAG differs from a real transportation network in that it is cycle-free, so the log-likelihood function is always computable regardless of the values of the choice parameters. In contrast, when it comes to route choice modeling, computing the log-likelihood may necessitate higher values for the choice parameters \citep{MaiFrejinger22}.

It can be observed that the composite alternatives may share a complex correlation structure due to inter-overlappings between subsets of alternatives. 
By bringing route choice models into the context, we further show that the nested RL \citep{MaiFosFre15}, which was developed to capture the correlation between path alternatives in a transportation network, can be conveniently used.

We evaluate our modeling approach and the proposed estimation algorithms with synthetic datasets of different sizes, and with two real-world datasets. The results demonstrate that 
our methods are highly scalable for handling large-scale settings and provide better in-sample and out-of-sample results compared to other RUM-based baselines. Moreover, our results based on real datasets also show that the nested RL model outperforms the RL model (on the DAGs) in terms of both in-sample and out-of-sample fits.

Our DAG-based representations are highly general and can incorporate other types of constraints and settings, such as customer budgets or specific constraints on the selected items. These constraints can be transformed into additional information that can be incorporated into nodes of the DAG. This may result in a large-sized DAG, but previous work has shown that recursive route choice models can handle networks of up to half a million nodes and arcs \citep{mai2021RL_STD,de2020RL_dynamic}, in terms of both estimation and prediction. All these imply the scalability and flexibility of our modeling approach.

\noindent
\textbf{Paper outline:}
Section \ref{sec:review} presents a literature review. Section \ref{sec:LMDC} presents a logit-based model for the multiple discrete choice problem. Section \ref{sec:DAGS} presents the DAG-based presentations  and their properties. In Sections \ref{sec:RLmodel}  and \ref{sec:NestedRL-model} we present the RL and nested RL model on the DAGs and show equivalence properties. Section \ref{sec:exp} provides numerical experiments and Section \ref{sec:concl} concludes. The appendix includes proofs that are not shown in the main part of the paper, and a discussion on an extension to a multiple discrete-count problem, i.e., individuals can select an elemental alternative multiple times.  

\noindent
\textbf{Notation:}
Boldface characters represent matrices (or vectors), and $a_i$ denotes the $i$-th element of vector $\ba$, if $\ba$ is indexable.  For any integer number $m>0$, let $[m]$ denote the set $\{1,2,\ldots,m\}$.

\section{Literature Review}\label{sec:review}
Numerous demand models have been developed to analyze people's multiple-choice behavior using a constrained utility maximization framework. These models often assume that a choice-maker would maximize an overall utility while satisfying the constraints they face. For instance, \cite{kim2002modeling} develop a direct utility model that allows for both choices of single goods, and joint choice of multiple goods. \cite{bhat2005multiple,bhat2008multiple} propose the Multiple Discrete-Continuous Extreme Value (MDCEV) model, which considers both package choice of multiple elemental alternatives, along with the continuous amount of consumption for the chosen elemental alternatives. The MDCEV model has been used across various fields due to the existence of closed-form choice probabilities \citep[see][for instance]{ma2019modeling,varghese2019multitasking,mouter2021contrasting}. These models assume a continuous decision space and rely on Karush-Kuhn-Tucker (KKT) conditions to ensure the optimality of observed demand. Several extensions have been proposed for incorporating
important factors of consumer decisions such as incorporating utility complementarity \citep{bhat2015allowing}, multiple constraints \citep{satomura2011multiple,castro2012accommodating} and non-linear pricing \citep{howell2016price}.

Despite many successes, it remains challenging for the above models to capture the discrete nature of people's behavior, which motivated \cite{bhat2022closed} to propose the Multiple Discrete-Count (MDC) model for multiple discrete-count choice modeling. The MDC model consists of two components: a total count model and an MDCEV model, where the discrete component corresponds to whether or not an individual chooses a specific elemental alternative over the total count, and the continuous component refers to the fractional split of investment in each of the consumed alternatives over the total count. While the MDC model has a closed-form solution and would be useful in multiple discrete choice applications, one possible drawback is that it involves many parameters, which could make it expensive to estimate. Additionally, the probability of zero counts involves enumerating many subsets of the goods, which can be intractable to handle exactly.

Several other structural models for multiple discrete demand have also been proposed. \cite{hendel1999estimating} proposed a multiple-discrete choice model to examine differentiated products demand and applied it to firm-level survey data on demand for personal computers. \cite{dube2004multiple} employed the same modeling framework to analyze household panel data of carbonated soft drinks. \cite{van2018simultaneous} built a model based on the assumption that a consumer's purchase decision maximizes variety while satisfying utility constraints. The solution to the constrained optimization problem is found by a sequential update procedure that reflects the discrete nature of consumer demand. \cite{lee2014modeling} extended models of constrained utility maximization by introducing demand indivisibility as an additional constraint, showing that ignoring the discrete nature of indivisible demand could result in biases in parameter estimates and policy implications. All the aforementioned models are based on constrained utility maximization, i.e., a choice-maker finds an exact optimal solution that maximizes his/her overall utility while satisfying potentially complex constraints. As mentioned earlier, this assumption might conflict with the importance of considering bounded rationality in decision-making and would be too strong because constrained utility maximization is not easy to solve. 

Our modeling approach is distinct from the aforementioned works in that we directly extend the RUM framework and classical single discrete choice models (such as the MNL) to account for multiple discrete behaviors without imposing a strong assumption on how choice-makers choose composite alternatives. We assume only that the choice-makers have a constraint on the range of the number of elemental alternatives they will select, noting that our framework is general to incorporate more complex constraints. Furthermore, by demonstrating the equivalence between a multiple discrete choice decision and a path choice in a network of choices, we provide an intuitive interpretation of how individuals make complex multiple choices.

As we bring route choice models, particularly link-based recursive models, into the context of multiple discrete choice analysis, it is relevant to discuss some of the current state-of-the-art models in the literature. Route choice modeling has been approached using both path-based and link-based recursive models. Path-based models, as reviewed by \cite{Prat09}, rely on path sampling, which may lead to parameter estimates that depend on the sampling method. In contrast, link-based recursive models are constructed using the dynamic discrete choice framework \citep{RUST1987}, and are essentially equivalent to discrete choice models based on all possible paths. Link-based recursive models have several advantages, including consistent estimation and ease of prediction, which have made them increasingly popular. The recursive logit (RL) model, proposed by \cite{FosgFrejKarl13}, was the first link-based recursive model developed. Since then, several recursive models have been introduced to handle correlations between path utilities \citep{MaiFosFre15,Mai_RNMEV,MaiBasFre15_DeC}, dynamic networks \citep{de2020RL_dynamic}, stochastic time-dependent networks \citep{mai2021RL_STD}, or discounted behavior \citep{oyama2017discounted}, with applications in areas such as traffic management \citep{BailComi08,Melo12} and network pricing \citep{zimmermann2021strategic}. For a comprehensive review, we suggest referring to \cite{zimmermann2020tutorial}.

\section{Logit-based Multiple Discrete Choice Model}\label{sec:LMDC}
Let $[m] = \{1,2,\ldots,m\}$ be the set of the elemental alternatives. It is assumed that each individual $n$ associated a utility $u_{ni}$ with each elemental choice alternative $i\in [m]$.
Under the RUM framework \citep{McFa81}, each utility $u_{ni}$ is the sum of two parts $u_{ni} = v_{ni} +\epsilon_{ni}$,  where $\epsilon_{ni}$ is a random term that cannot be observed by the analyst, and the deterministic term $v_{ni}$ can include attributes of the alternative as well as socio-economic
characteristics of the individual. In general, a linear-in-parameters formula is used, i.e, $v_{ni}(\bbt) =\bbt^\transpose x_{ni}$, where $\bbt$ is a vector of parameters to be estimated and $\bx_{ni}$ is a vector of attributes of alternative $i$ as observed by individual $n$.
In the classical single-choice setting, the RUM framework assumes that individuals make a choice decision by maximizing their random utilities, yielding the choice probabilities
\[
P(i|[m]) = P(u_{ni}\geq u_{nj};~\forall j\in [m]) = P(v_{ni}(\bbt) + \epsilon_{ni}  \geq v_{nj}(\bbt)+\epsilon_{ni};~\forall j\in [m]) 
\]
In the multiple discrete choice setting, each individual will choose a subset of elemental alternatives, instead one single choice alternative $j\in [m]$. So, each composite alternative now is a subset $S$ of $[m]$. To extend the single RUM model to address the multiple discrete choice behavior, for each composite alternative $S\subseteq [m]$, let $v^n(S|\bbt)$ be its deterministic utility. We assume that $v^n(S|\bbt)$ can be written as a sum of the  utilities  of  the elemental alternatives in $S$, i.e., $v^n(S|\bbt) = \sum_{i\in S} v_{ni}(\bbt)$. Under the RUM framework, we now assume that each individual selects a composite alternative by maximizing its random utility $u^n(S|\bbt) = v^n(S|\bbt) + \epsilon_{nS}$, from a set of composite alternatives that she finds appealing. We further assume that individual $n$ will not consider all the possible subsets of $[m]$. Instead, she is only interested in subsets whose sizes lie in the interval $[L_n,U_n]$, where $L_n,U_n \in [m]$ and $L_n\leq U_n$. These bounds represent the minimum and maximum allowable sizes for a composite alternative to be considered by the choice-maker. Let us denote $\Omega^{[L_n,U_n]}$ as the consideration set of individual $n$, $\Omega^{[L_n,U_n]} = \{S\subseteq [m]|~ |S| \in [L_n,U_n]\}$, the choice probability of  a composite alternative $S\in \Omega^{[L_n,U_n]} $ can written as
\[
P(S|~\Omega^{[L_n,U_n]},\bbt) = P(v_n(S|\bbt)+\epsilon_{nS} \geq v(S'|\bbt) +\epsilon_{nS'};~\forall S'\in \Omega^{[L_n,U_n]})
\]
If the random terms $\epsilon_{nS}$ are i.i.d Extreme Value Type  I (i.e., the choice model over 
 $\Omega^{[L_n,U_n]}$ is the MNL model), then the choice probabilities become 
 \begin{equation}     
 \label{eq:multi-choice-RUM}
P(S|~\Omega^{[L_n,U_n]}) =\frac{\exp(v(S|\bbt))}{\sum_{S'\in \Omega^{[L_n,U_n]}} \exp(v(S'|\bbt))} = \frac{\exp\left(\sum_{i\in S}v_{ni}(\bbt)\right)}{\sum_{S'\in \Omega^{[L_n,U_n]}} \exp\left(\sum_{i\in S'}v_{ni}(\bbt)\right)}  
 \end{equation}
We refer to this choice model as the logit-based multiple discrete choice model (LMDC).  The model will collapse to the single-choice MNL model when $L=U=1$.
The number of elements in $\Omega^{[L_n,U_n]}$ is $\sum_{t=L_n}^{U_n} {m \choose t} $, which  is generally exponential in $m$. So, it is practically not possible to compute $P(S|~\Omega^{[L_n,U_n]})$ by enumerating all the elements in the set. This is the main challenge arising from the use of \eqref{eq:multi-choice-RUM}. The complex overlapping structure between the composite alternatives within $\Omega^{[L_n,U_n]}$ also presents a challenge in capturing the correlation using conventional single choice models such as the classical nested or cross-nested logit model \citep{BenA73,BenABier99a}. We will show later  that such correlations can be captured naturally using the nested recursive route choice model \citep{MaiFosFre15}.

Now, given a set of observations $\cD = \{(S_n,L_n,U_n),~ n\in [N]\}$, where each observation comprises an observed composite alternative and two integral values representing the minimum and maximum sizes of the subsets being considered by the corresponding choice-maker. The log-likelihood function of the observation data can be defined as 
\begin{equation}\label{eq:LL-func}
    \cL(\cD|\bbt)  = \sum_{n\in [N]} \ln P(S_n|~\Omega^{[L_n,U_n]}).  
\end{equation}
The model parameters $\bbt$ can be estimated via maximum likelihood estimation (MLE), i.e. solving $\max_{\bbt} \cL(\bbt)$. It is well-known that, in the context of single choices and the choice model is MNL, if the utility functions are linear-in-parameters, then 
the log-likelihood function is concave in the model parameters. Despite having a more complex structure, it is possible to extend this result to show that the log-likelihood function in \eqref{eq:LL-func} also manifests concavity in $\bbt$. 
\begin{proposition}
If $v_{ni}(\bbt)$ are linear in $\bbt$, $\forall n\in [N],~i\in [m]$ and the choice model over the composite alternatives is the MNL,  then the function $\cL(\bbt)$ is concave in $\bbt$.
\end{proposition}
\proof{}
We simply write the log-likelihood function under the MNL model as
\begin{align}
\cL(\cD|\bbt) &=  \sum_{n\in [N]} \ln P(S_n|~\Omega^{[L_n,U_n]}) \nonumber \\
&=\sum_{n\in [N]} v(S_n|\bbt) -\sum_{n\in [N]} \ln \left( \sum_{S'\in \Omega^{[L_n,U_n]}} \exp\left(v(S'|\bbt)\right)\right).\label{eq:proof-concave-mnl-eq1}
\end{align}
Since $v_{ni}(\bbt)$ is linear in $\bbt$, the first part of \eqref{eq:proof-concave-mnl-eq1} is linear in $\bbt$  and the second part of \eqref{eq:proof-concave-mnl-eq1} is a log-sum-exp function of linear functions, which is also concave \citep{boyd2004convex}. So, $\cL(\bbt)$ is also concave as desired. 
\endproof

The concavity can guarantee global optimums when solving the MLE problem. This property generally does not hold if $v_{ni}(\bbt)$ is not linear in $\bbt$, or the choice model is not the MNL.

The computation of the log-likelihood function $\cL(\bbt)$ involves handling sets of exponentially many elements $\Omega^{[L_n,U_n]}$. Therefore, it would be impractical to compute it directly. We will deal with this challenge in the next section by converting the choices over composite alternatives into a path choice model in a cycle-free network.

\section{DAG-based Representations}\label{sec:DAGS}
In this section, we present two DAG-based representations for the multiple-choice problem.
Our idea is to build a cycle-free directed graph $\cG=(\cN,\cA)$ where $\cN$ is the set of nodes and $\cA$ is the set of arcs. Each node in $\cN$  will contain information about whether an elemental alternative is chosen or not, and the number of chosen elemental alternatives. We construct the DAGs in such a way that there is a one-to-one mapping between a composite alternative in the multiple-choice problem and a path choice in the DAG. It then leads to the result that a recursive route choice model on the DAG will be equivalent to the LMDC model described in the previous section. The DAG representations will be introduced below, with the subscript $n$ used to index the choice-maker omitted for notational simplicity.


\subsection{Binary-Choice DAG}

The first DAG representation draws inspiration from the notion that, when confronted with a set of discrete choices, an individual may consider each option one by one and decide whether to choose it or not, keeping in mind the current count of the already chosen items. When the number of selected elemental alternatives reaches its upper bound,  the choice-maker will stop the selection process. In order to represent this as a DAG, a graph will be constructed with nodes arranged in $m+2$ tiers. The first tier will comprise a single node that denotes the sole origin of the graph, while the final tier will likewise comprise a single node representing the only destination of the DAG. Each intermediate tier between the origin and destination will provide details pertaining to one of the $m$ elemental alternatives. Each node at an intermediate tier between the origin and destination will encompass details regarding whether the corresponding elemental alternative is chosen or not, and the number of selected alternatives during the choice process. The DAG representation described above is referred to as the \textit{Binary-Choice DAG} (or \textit{BiC} for short). The name \textit{Binary-Choice} reflects the fact that two options are presented for each elemental alternative -- either being chosen or rejected.
We illustrate this DAG in \autoref{fig:bin-graph} below.

\begin{figure}[htb] 
\centering
    \includegraphics[width=0.8\linewidth]{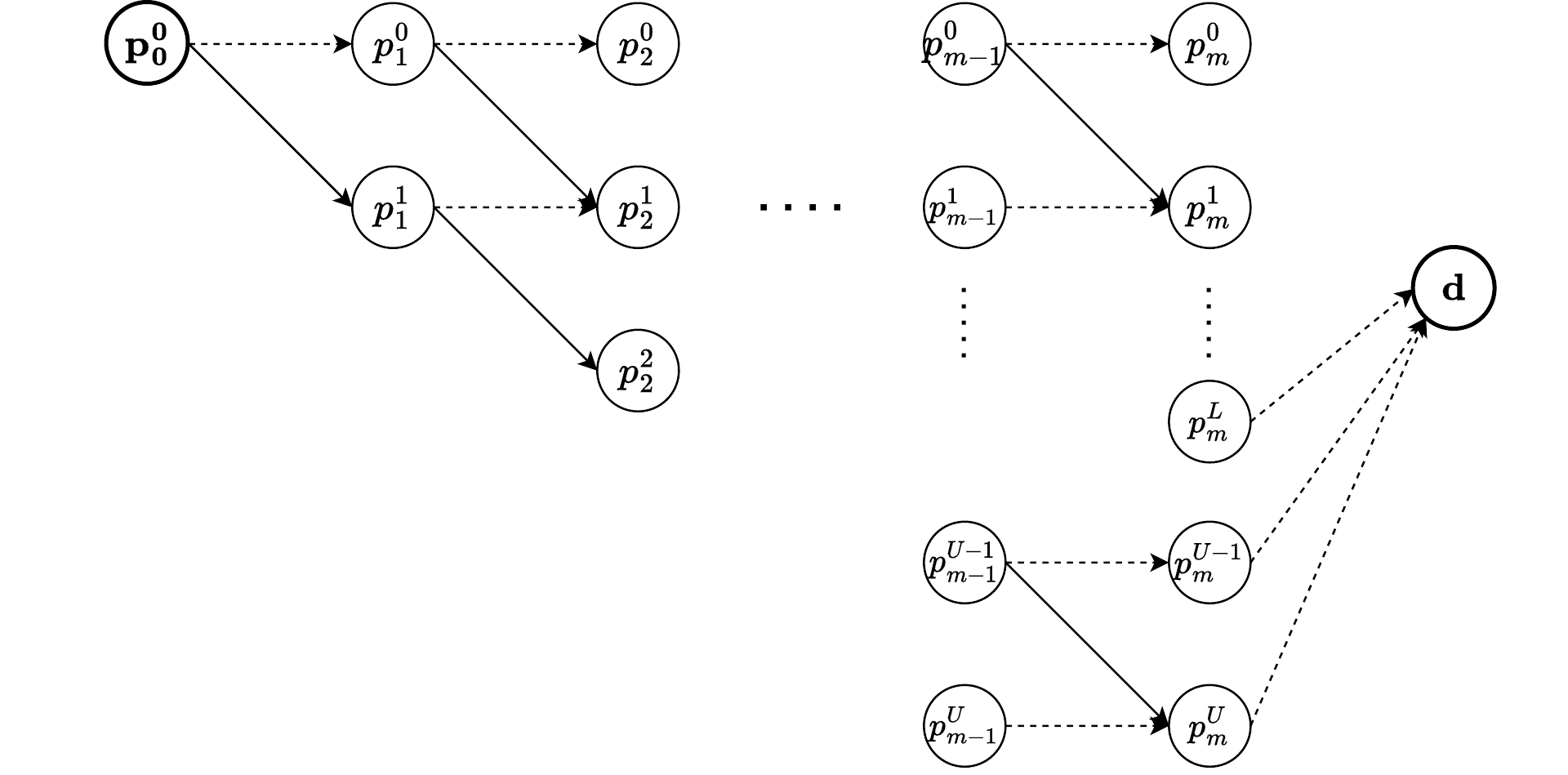} 
    \caption{Binary-Choice DAG.} 
    \label{fig:bin-graph} 
\end{figure}


We create arcs in $\cA$ to connect nodes in the $k$-th tier to $(k+1)$-th tier, for $k=0,1,\ldots,m$. Each arc of the DAG will represent a binary decision -- choosing or not choosing the next elemental alternative. For instance, from $p^0_0$, there are two arcs connecting it with $p^0_1$  and $p^1_1$, where the arc $(p^0_0,p^0_1)$ represents the option of not selecting alternative 1,  and the arc $(p^0_0,p^1_1)$ represents the option of selecting that alternative.   All nodes in the $m$-th tier (the second last tier) will connect to the destination $d$, which is an absorbing node. For any node of form $p^{U}_k$, we create an arc directly connecting it to the destination, as the choice process already reaches the maximum number of elemental alternatives allowed and must be terminated. Furthermore, for a node $p^k_m$ where $k\leq L$, we do not connect it with any other node of the DAG. In this case, the choice process reaches the last stage but has not gotten a sufficient number of choice alternatives to form a feasible composite choice. The following proposition shows that there is a one-to-one mapping between a path in the BiC DAG with a composite alternative.
\begin{proposition}\label{prop:map-BiC}
    There is a one-to-one mapping between any composite alternative $S\in \Omega^{[L,U]}$ and a path from $p^0_0$ to $d$ in the BiC DAG. 
\end{proposition}
\proof
   For any composite alternative $S \in  \Omega^{[L,U]}$, we then create a path $\{p^{j_0}_0,\ldots,p^{jm}_m,d\}$ from $p^0_0$ to $d$ such that, for any $i\in [m]$, we follows the arc $(p^{j_{i-1}}_{i-1}, p^{j_{i-1}}_{i})$ if $i\notin S$ (i.e., alternative $i$ is not chosen), and follows the arc  $(p^{j_{i-1}}_{i-1}, p^{j_{i-1}+1}_{i})$ if $i\in S$ ($i$ is chosen). Since  $S\in [L,U]$, such a path always exists and can be uniquely created. For the  opposite way,  let $\{p^{j_0}_0,p^{j_1}_1,p^{j_2}_2,\ldots,p^{jm}_m,d\}$ be a path connecting $p^0_0$ and $d$. We can create a composite alternative  $S$ by adding item $i$ to $S$, for any $i\in [m]$, if arc $(p^{j_{i-1}}_{i-1}, p^{j_{i}}_{i})$ corresponds to a ``selecting'' option, i.e., $j_{i} = j_{i-1}+1$. The way we created the DAG implies that the size of $S$ should be in the range $[L,U]$, thus $S \in  \Omega^{[L,U]}$. We complete the proof. 
\endproof

Proposition  \ref{prop:size-BiC} below gives the exact numbers of nodes and arcs of the BiC DAG. 
\begin{proposition}\label{prop:size-BiC}
    The number of nodes and arcs in the BiC DAG are 
    \begin{align*}
        |\cN| &= m + mU - \frac{U^2}{2} + \frac{U}{2} + 2 \\
        |\cA| &= 2m + 2mU - U^2 - L + 1. 
    \end{align*}
\end{proposition}




\subsection{Multiple-Choice DAG}

\begin{figure}[htb] 
\centering
    \includegraphics[width=0.8\linewidth]{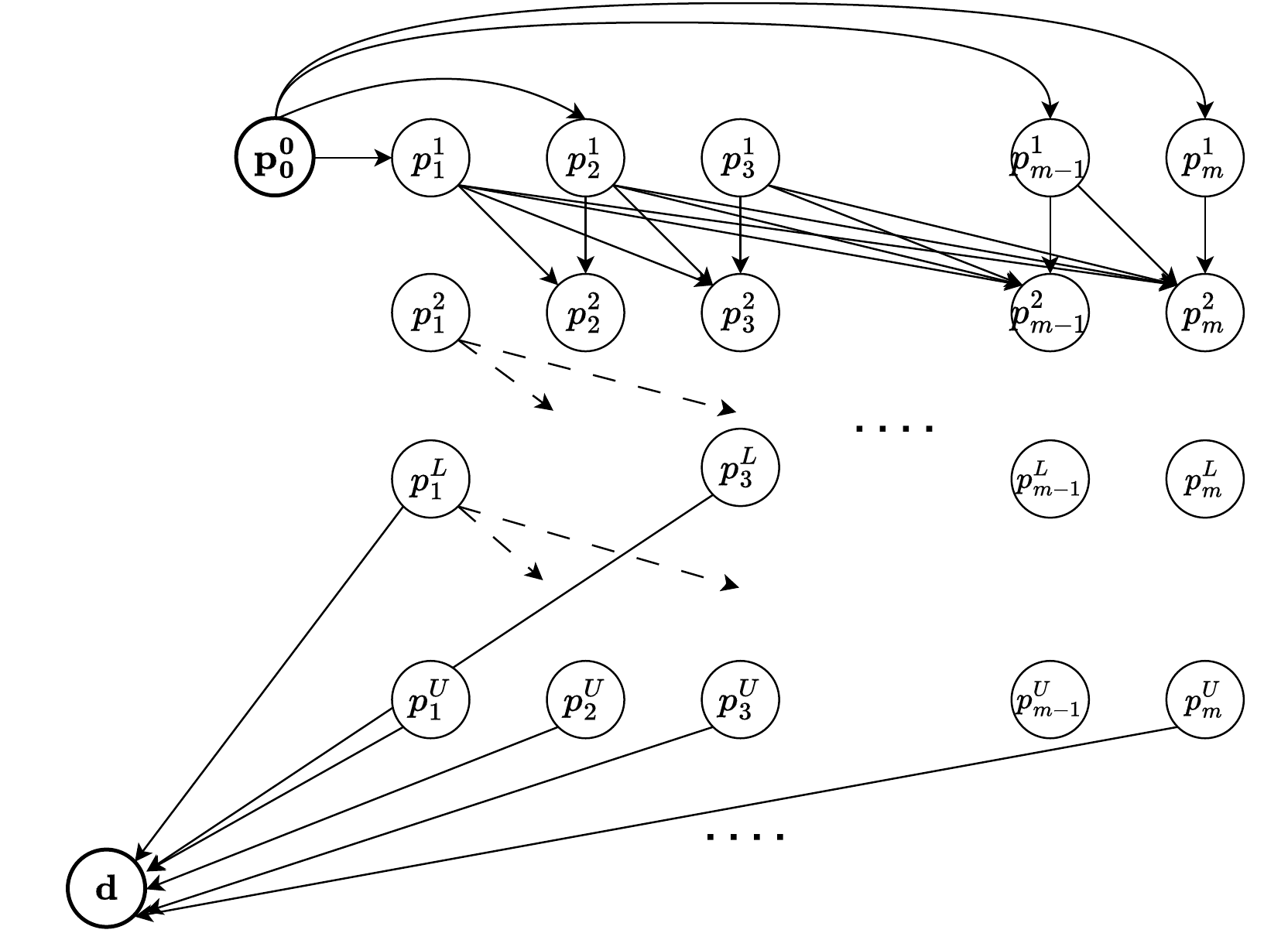} 
    \caption{Multi-Choice DAG.} 
    \label{fig:choice-graph} 
\end{figure}

The second DAG representation takes a different perspective on the multiple-choice behavior. Rather than assuming that individuals consider each option one by one and decide whether to ``select'' or ``reject'' it, this modeling approach presumes that the individuals immediately navigate to an alternative of interest, choosing them one by one. We represent this choice process by the DAG in \autoref{fig:choice-graph} where nodes are also arranged into $m+2$ tiers, each corresponds to an elemental alternative indexed from 1 to $m$. There is also an original node, denoted as $p^0_0$, and a destination $d$. Each node of the DAG, except the destination, will be connected to nodes in the subsequent tiers, reflecting the fact that individuals can directly navigate to  the next alternatives that they are interested. Each node in the DAG (except the destination) is denoted as $p^c_j$, where $j\in [m]$ represents the index of the corresponding tier, and the superscript $c$ represents the total number of selected elemental alternatives.
As a result, there is no node with $c>U$. 
For each node $p^c_j$ of the DAG, we create arcs connecting it with nodes $\{p^{c+1}_{j+1},\ldots,p^{c+1}_{m}\}$, reflecting the fact that individuals can select any subsequent alternatives in $\{j+1,\ldots,m\}$, and the number of selected items is increased by 1. For nodes of the form $p^c_k$, for any $k\in [m]$ and $L\leq c\leq U$, we directly connect them to the destination, as the number of selected items is already in its allowed range $[L, U]$.  We call this DAG as\textit{ Multi-Choice DAG} (or MuC DAG for short). The name \textit{Multi-Choice} is used because, in contrast to the BiC DAG where each node has two arcs representing the options of selecting or rejecting, there are several arcs connecting each node to other nodes in the {MuC} DAG.

Similar to the BiC DAG, we can also show that there is a one-to-one mapping between any composite alternative  $S$ and a path from $p^0_0$ to $d$ in the MuC DAG.
\begin{proposition}\label{prop:map-MuC}
There is a one-to-one mapping between any composite alternative $S\in \Omega^{[L,U]}$ and a path from $p^0_0$ to $d$ in the \textit{MuC} DAG. 
\end{proposition}
\proof{}
The proof can be easily verified given the way the DAG is constructed. Let $S = \{j_1,j_2,\ldots,j_K\}$ be a feasible composite alternative, where $K\in [L,U]$ and $j_1<j_2<\ldots <j_K$. We map $S$ with the  path  $(p^0_0, p^1_{j_1}, p^2_{j_2}, \ldots, p^K_{j_K}, d)$, where each arc  $(p^k_{j_k}, p^{k+1}_{j_{k+1}})$, $k=1,\ldots,K - 1$, corresponds to the selection of the elemental alternative $j_{k+1}$ after $j_k$. It can be seen that  $(p^0_0, p^1_{j_1}, p^2_{j_2}, \ldots, p^K_{j_K}, d)$ is a feasible path from $p^0_0$ to $d$, and the mapping is unique. For the opposite direction, given a feasible path $(p^0_0,p^1_{j_1},...,p^K_{j_K})$, where $j_k\in [m]$, $k=\in [K]$, and $j_1<j_2<\ldots,j_K$,  we can uniquely map it with the subset $\{j_1,\ldots,j_K\}$. The way we constructed the MuC DAG implies that this subset should belong to $\Omega^{[L,U]}$, which completes the proof.
\endproof

The following proposition gives the exact numbers of nodes and arcs of the MuC DAG. 
\begin{proposition}\label{prop:number-node-MuC}
The numbers of nodes and arcs in the MuC DAG are 
\begin{align*}
    |\cN| &= mU - \frac{U^2}{2} + \frac{U}{2} + 2 \\
    |\cA| &= 
    \begin{cases}
    \frac{m^2U}{2} - \frac{mU^2}{2} + mU + 2m + \frac{U^3}{6} - \frac{U^2}{2} + \frac{U}{3} + 2,   & L = 0 \\
    \frac{m^2U}{2} - \frac{mU^2}{2} - mL + mU + 2m + \frac{L^2}{2} - \frac{3L}{2} + \frac{U^3}{6} - \frac{U^2}{2} + \frac{U}{3} + 1,   & L > 0
    \end{cases}.
\end{align*}
\end{proposition}

Compared to the size of the BiC DAG considered in the previous section, we can see that the numbers of nodes in the BiC and MuC DAG are quite similar -- both are in $\cO(mU)$, while the MuC DAG has significantly more arcs; it has $\cO(m^2U)$ arcs, compared to $\cO(mU)$ arcs in the BiC DAG.





\section{Recursive Logit Model on the DAGs}\label{sec:RLmodel}

In this section, we first present an RL model \citep{FosgFrejKarl13} applied to the DAGs previously introduced. We then show that the choice probabilities over the composite alternatives are the same as the path choice probabilities yielded by the RL model, enabling us to utilize the estimation methods developed in prior work \citep{FosgFrejKarl13,MaiFosFre15} to efficiently estimate the LMDC model. 

\subsection{Recursive Logit Model on the DAGs}
Consider a DAG $G=(\cN, \cA)$ where $\cN$ is the set of nodes and $\cA$ is the set of arcs. For each node $k\in \cN$, let $N(k)$ be the set of nodes that can be reached directly from $k$, i.e., $N(k) = \{a|~ (k,a)\in \cA\}$.  
 A path in the DAG is a sequence of nodes $(k_0, k_1,..., k_L)$ in which $k_0 = p^0_0$ (the origin of the DAG), $k_L = d$ (the destination of the DAG), and $k_{i+1}\in N(k_i),~ \forall i = 0,\ldots,L-1$. It can be seen that, for both BiC and MuC DAGs, the length of any path is $m+2$. We associate each arc  $(k,a)\in \cA$ with a utility $v(a|k)$. Specifically, for the BiC DAG, we set
 \begin{equation}\label{eq:v-BiC}
\begin{cases}
 v^{\BiC}(p^c_{j+1}|p^c_j) = 0,~c\in \{0,\ldots,U\},~j \in \{0,...,m\}    \\
 v^{\BiC}(p^{c+1}_{j+1}|p^{c}_{j}) = v_j ~c\in \{0,\ldots,U-1\},~j \in \{0,...,m\}    \\
 v^{\BiC}(k|d) = 0, ~\forall k\in \cN, d\in N(k), 
\end{cases}
 \end{equation}
 noting that all the utilities $v^{\BiC}(a|k)$ are functions of the choice parameters $\bbt$.
Intuitively, for each arc $(k,a)$ of the BiC DAG, we set $v(a|k) = 0$ if the arc $(k,a)$ represents a ``non-selecting'' option. Otherwise, we set $v(a|k)$ to the utility of the elemental alternative that is associated with $a$. In particular, we assign a zero utility for arcs that connects nodes to the destination $d$.  

For the MuC DAG, we set the arc utilities as follows
 \begin{equation}\label{eq:v-MuC}
\begin{cases}
 v^{\MuC}(p^{c+1}_{j+1}|p^c_j) = v_j,~c\in \{0,\ldots,U-1\},~j \in \{0,...,m\}    \\
 v^{\MuC}(k|d) = 0, ~\forall k\in \cN, d\in N(k). 
\end{cases}
\end{equation}
That is, for any arc $(k,a)\in \cA$, we set $v^{\MuC}(a|k)$ to the utility of the elemental alternative that is associated with $a$. All the arcs that connect nodes to the destination will be set to zero utilities. In the following, we use the notation $v^{\BiC}(.)$ to present the RL model. The RL model applied to the MuC DAG can be formulated in the same way.

In the context of the RL model, a path choice is modeled as a sequence of arc choices. It is assumed that, at each node 
$k\in \cN$, $k\neq d$,  the traveler chooses the next node $a$ from the set of possible outgoing nodes $A(k)$ by maximizing the sum of the arc utility $v^{\BiC}(a|k)$ and an expected downstream utility (or value function) $V^{\BiC}(a)$ from node $a$ to the destination. This value function is defined recursively as  
\begin{equation*}
    V^{\BiC}(k)  = 
    \begin{cases}
    0,   & k = d \\
    -\infty & \text{ if } N(k)= \emptyset \\
    E_{\epsilon} \left[\max_{a\in A(k)} \{v^{\BiC}(a|k) + \epsilon(a) + V^{\BiC}(a) \}\right],   & \forall k\in\cN\backslash \{d\}
    \end{cases}
\end{equation*}
where $\epsilon(a)$  are i.i.d Extreme Value Type I
 with scale $\mu>0$  and are independent of everything in the network. The value function can be rewritten as
\begin{equation}\label{eq:value}
    V^{\BiC}(k)  = 
    \begin{cases}
    0,   & k = d \\
     -\infty & \text{ if } N(k)= \emptyset \\
    \mu \ln{\Bigg(\sum_{a\in N(k)} \exp{\left( \frac{1}{\mu} (v^{\BiC}(a|k) + V^{\BiC}(a)) \right)}\Bigg)},   & \forall k\in\cN\backslash\{d\}.
    \end{cases}
\end{equation}
Moreover, according to the closed form of the logit model, the probability of choosing node $a$  from node $k$ is
\[
\begin{aligned}    
P^{\BiC}(a|k) &= \frac{\exp{\left( \frac{1}{\mu} (v^{\BiC}(a|k) + V^{\BiC}(a)) \right)}}{\sum_{a'\in N(k)} \exp{\left( \frac{1}{\mu} (v^{\BiC}(a'|k) + V^{\BiC}(a')) \right)}}\\
&=  \frac{\exp{\left( \frac{1}{\mu} (v^{\BiC}(a|k) + V^{\BiC}(a)) \right)}}{ \exp{\left( \frac{1}{\mu} (V^{\BiC}(k)  \right)}}  ,~\forall (k,a)\in \cA.
\end{aligned}
\]
The probability of a path $\sigma = \{k_0,\ldots,k_{m+1}\}$ in the DAG can be computed as the product of the arc probabilities
\begin{equation}
\label{eq:choice-prob-bic}
P^{\BiC}(\sigma|\bbt) = \prod_{i=0}^{m} P^{\BiC}(k_{i+1} | k_i) = \frac{\exp{\left( \frac{1}{\mu} v^{\BiC}(\sigma) \right)}}{\exp{\left( \frac{1}{\mu} V^{\BiC}(k_0) \right)}} = \frac{\exp{\left( \frac{1}{\mu} v^{\BiC}(\sigma) \right)}}{\sum_{\sigma' \in \Omega^{\BiC}}\exp{\left( \frac{1}{\mu} v^{\BiC}(\sigma') \right)}}  
\end{equation}
where $v^{\BiC}(\sigma) = \sum_{i=0}^{m} v^{\BiC}(k_{i+1}|k_i)$ and $\Omega^{\BiC}$ is the set of all the paths from $p^0_0$ to $d$. Given a set of path observations $\cD = \{\sigma_1,\ldots,\sigma_T\}$, the RL can be estimated by maximizing  the log-likelihood function
\begin{equation}\label{eq:LL-func-graph}
    \mathcal{L}^{\BiC}(\cD|\bbt) = \frac{1}{\mu}\sum_{t\in [T]} \left(v^{\BiC}(\sigma_t) - V^{\BiC,t}(k_0^t)\right)  
\end{equation}
where $k_0^t$ is the first node of path $\sigma_t$, $t\in [T]$. Here, we note that the value function $V^{\BiC,t}(\cdot)$ has superscript $t$ which reflects the fact that it could be observation dependent, as 
 different individuals may have different lower and upper bounds on the number of selected alternatives (i.e., $L$ and $U$), leading to different DAG structures, and different value functions.

\subsection{Equivalence Properties}\label{subsec:model-equal}
Having introduced the RL model on the MuC and BiC DAG, we now connect the RL to the LMDC model presented in Section \ref{sec:LMDC}.  Propositions \ref{prop:map-BiC} and \ref{prop:map-MuC} above show that there are one-to-one mappings between any subset $S\in \Omega^{[L,U]}$ to a path in the BiC or Muc DAG. 
To facilitate the later exposition, for each subset $S\in \Omega^{[L,U]}$, let $\sigma^{\BiC}(S)$ and $\sigma^{\MuC}(S)$ be the two paths in the BiC and MuC DAGs, respectively, that maps with $S$.
Theorem \ref{th:equiv-prob} below states our main result saying that the choice probability $P(S|~\Omega^{[L,U]})$ are equal to the probabilities of the corresponding paths in the BiC and MuC DAGs, given by the RL model.  

\begin{theorem}\label{th:equiv-prob}
For any $S\in \Omega^{[L,U]}$ we have
\[
P(S
|~\Omega^{[L,U]}) = P^{\BiC}(\sigma^{\BiC}(S)) =   P^{\MuC}(\sigma^{\MuC}(S)). 
\]
\end{theorem}
\proof{}
We first prove the claim for the BiC DAG. For a subset $S\in \Omega^{[L,U]}$, Proposition \ref{prop:map-BiC} tells us that $\sigma^{\BiC}(S) = \{p^{j_0}_0,\ldots,p^{j_m}_m,d\}$ such that $p^{j_0}_0 = p^0_0$ and for any $i\in[m]$, if $i\in S$ then $j_{i} = j_{i-1}+1$ (i.e., alternative $i$ is selected), and if  $i\notin S$ then $j_{i} = j_{i-1}$ (i.e., alternative $i$ is not selected). The total arc utility of $\sigma^{\BiC}(S)$ becomes $\sum_{i=0}^m v^{\BiC}(p^{j_{i+1}}_{i+1}|p^{j_{i}}_{i})$. Moreover, the definition of $v^{\BiC}(\cdot)$ in \eqref{eq:v-BiC} implies that $v^{\BiC}(p^{j_{i+1}}_{i+1}|p^{j_{i}}_{i}) = 0$ if $j_{i+1} = j_i$ and $v^{\BiC}(p^{j_{i+1}}_{i+1}|p^{j_{i}}_{i}) = v_i$ if $j_{i+1} = j_i+1$. Combine this with the above, we have $\sum_{i=0}^m v^{\BiC}(p^{j_{i+1}}_{i+1}|p^{j_{i}}_{i})  = \sum_{i\in S} v_i$, or $v(S) = v^{\BiC}(\sigma^{\BiC}(S))$.
Furthermore,  from \eqref{eq:multi-choice-RUM} and \eqref{eq:choice-prob-bic},  the choice probabilities of subset $S$ and path $\sigma^{\BiC}(S)$ can be computed as
\begin{align}
    P(S
|~\Omega^{[L,U]}) &=\frac{\exp\left(\frac{1}{\mu} v(S)\right)}{\sum_{S'\in \Omega^{[L,U]}}\exp\left(\frac{1}{\mu} v(S')\right)} \nonumber \\
 P^{\BiC}(\sigma^{\BiC}(S)) &= \frac{\exp{\left( \frac{1}{\mu} v^{\BiC}(\sigma^{\BiC}(S)) \right)}}{\sum_{\sigma' \in \Omega^{\BiC}}\exp{\left( \frac{1}{\mu} v^{\BiC}(\sigma') \right)}}. \nonumber
\end{align}
This directly implies that $ P(S
|~\Omega^{[L,U]}) =  P^{\BiC}(\sigma^{\BiC}(S))$ for any $S
\in \Omega^{[L,U]}$.

In a similar way, for the MuC DAG, given a subset $S = \{j_1,j_2,\ldots,j_K\} \in \Omega^{[L,U]}$ , where $K\in [L,U]$ and $j_1<j_2<\ldots <j_K$. Proposition  \ref{prop:map-MuC} tells us that $S$ can be uniquely mapped to path  $\sigma^{\MuC}(S) = (p^0_0, p^1_{j_1}, p^2_{j_2}, \ldots, p^K_{j_K}, d)$, where each arc  $(p^k_{j_k}, p^{k+1}_{j_{k+1}})$, $k=1,\ldots,K - 1$, corresponds to the selection of the elemental alternative $j_{k+1}$ after $j_k$. The accumulated utility of $\sigma^{\MuC}(S)$ becomes 
$$v^{\MuC}(\sigma^{\MuC}(S))  = \sum_{k=0}^{K-1}v^{\MuC}(p^{k+1}_{j_{k+1}}|p^{k}_{j_{k}}) \stackrel{(a)}{=}\sum_{k=0}^{K-1} v_{j_{k+1}} = \sum_{i\in S} v_i,
$$ 
where $(a)$ is due to the definition of $v^{\MuC}(\cdot)$ in \eqref{eq:v-MuC}. We then have $v^{\MuC}(\sigma^{\MuC}(S)) = v(S)$. Similar to the case of the BiC DAG,  the choice probabilities should be the same, i.e.,  $ P(S
|~\Omega^{[L,U]}) =  P^{\MuC}(\sigma^{\MuC}(S))$, which completes the proof. 
\endproof

Let $\cD = \{S_1,\ldots,S_N\}$ be a set of observed subsets and $\cD^\BiC = \{\sigma^{\BiC}(S_1),\ldots,\sigma^{\BiC}(S_N)\}$ and $\cD^{\MuC} = \{\sigma^{\MuC}(S_1),\ldots,\sigma^{\MuC}(S_N)\}$ be the sets of the mapped paths in the BiC DAG and MuC DAG, respectively. Theorem \ref{th:equiv-prob} directly implies that the log-likelihood of  $\cD$ is also equal to the log-likelihood functions of $\cD^{\BiC}$ and $\cD^{\MuC}$ defined based on the BiC and MuC DAG, respectively. As a result, the estimation of the LMDC model can be done by estimating the RL with the $\cD^{\BiC}$ or $\cD^{\MuC}$. We state this result in the following Corollary.
\begin{corollary}
    For any set of subset observations $\cD = \{S_1,\ldots,S_N\}$, we have
    \[
    \cL(\cD|\bbt) = \cL^{\BiC}(\cD^{\BiC}|\bbt) = \cL^{\MuC}(\cD^{\MuC}|\bbt).  
    \]
\end{corollary}

\subsection{Illustrative Example}\label{subsec:example}
We give a simple example to illustrate the equivalence between the LMDC model introduced in Section \ref{sec:LMDC} and the RL model on the DAGs. We take a simple instance with $m=3$ (there are 3 elemental alternatives) and the number of alternatives to be selected is in the range of $[1, 2]$ (i.e., $L =1,U=2$). The elemental utilities and the choice probabilities of the composite alternatives are provided in  \autoref{tab:example}. The probability of each subset is calculated using \eqref{eq:multi-choice-RUM}.

\begin{table}[htb]
    \centering
    \begin{tabular}{ll}
        \hline
        Item $s_j$  & Utility $v_{s_j}$ \\ \hline
        $s_1$       & -1 \\
        $s_2$       & -1.5 \\
        $s_3$       & -2 \\ \hline
    \end{tabular}
    \quad
    \begin{tabular}{llll}
         \hline
         No.    & Subset         & Subset utility   & $P(S_n)$ \\ \hline
         $S_1$  & $\{s_1\}$      & -1         & 0.414 \\
         $S_2$  & $\{s_2\}$      & -1.5         & 0.251 \\
         $S_3$  & $\{s_3\}$      & -2         & 0.152 \\
         $S_4$  & $\{s_1, s_2\}$ & -2.5         & 0.093 \\
         $S_5$  & $\{s_1, s_3\}$ & -3         & 0.056 \\
         $S_6$  & $\{s_2, s_3\}$ & -3.5         & 0.034 \\ \hline
    \end{tabular}
    \caption{LMDC model  with $m=3,\,L = 1,\,U = 2$}
    \label{tab:example}
\end{table}

\begin{figure}[htb]
    \centering
    \includegraphics[width=0.6\linewidth]{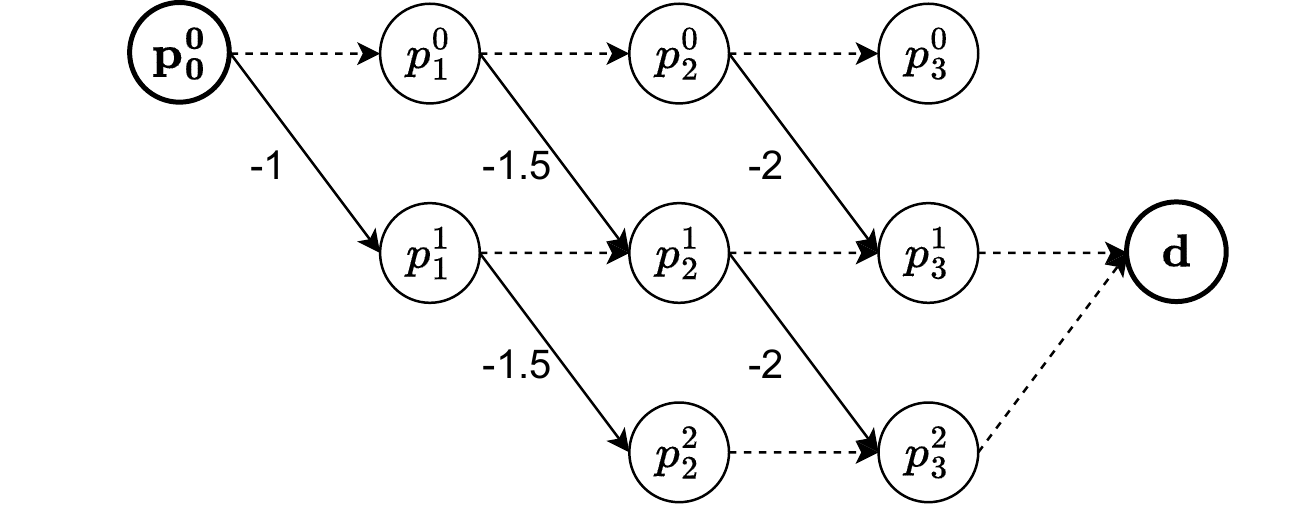}
    \caption{BiC DAG representation}
    \label{fig:bin-example}
\end{figure}

The corresponding BiC DAG for this example is illustrated in \autoref{fig:bin-example}, noting that dashed arcs represent "not-selecting" options and have zero utilities. In  \autoref{tab:bin-prob} we report the corresponding value function and path probabilities given by the RL model. It is easy to see the matching path $\sigma^n$ has the same probability as that of $S^n$ reported in \autoref{tab:example}.

\begin{figure}[htb]
    \centering    \includegraphics[width=0.6\linewidth]{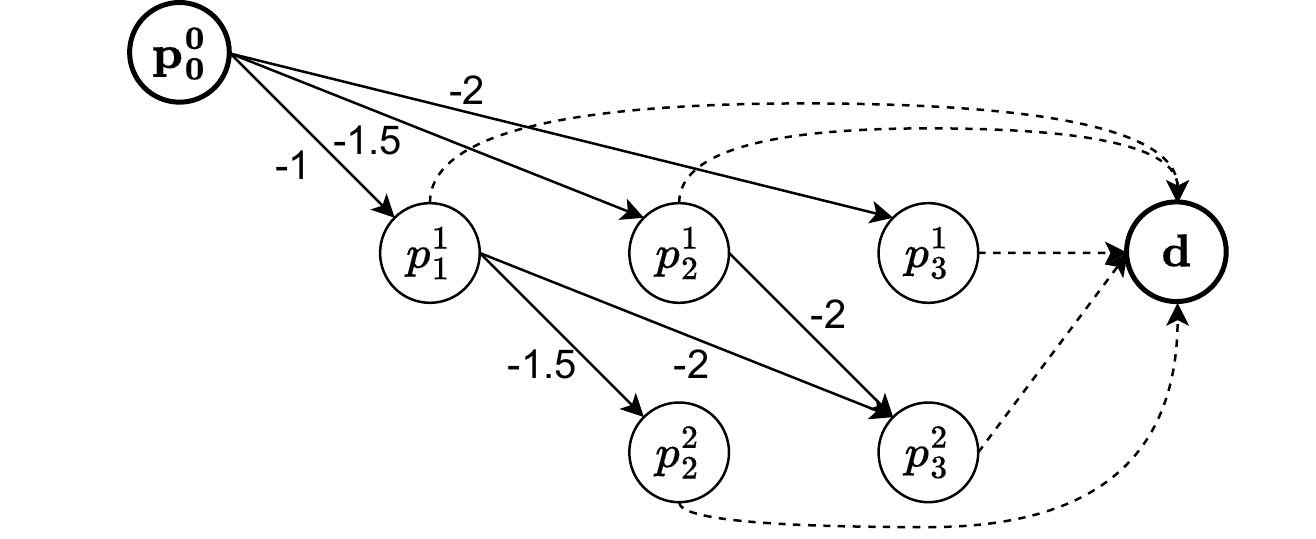}
    \caption{MuC DAG representation}
    \label{fig:multi-example}
\end{figure}

\begin{table}[htb]
    \centering
    \begin{tabular}{ll}
        \hline
        Node    & $V^{\BiC}(\cdot)$    \\ \hline
        $p_0^0$     & -0.118 \\
        $p_1^0$     & -0.945 \\
        $p_2^0$     & -2 \\
        $p_3^0$     & -$\infty$ \\
        $p_1^1$     & 0.306 \\
        $p_2^1$     & 0.127 \\
        $p_3^1$     & 0 \\
        $p_2^2$     & 0 \\
        $p_3^2$     & 0 \\
        $d$         & 0 \\
    \end{tabular}
    \quad
    \begin{tabular}{l|l|l|l}
No.        & Path  & Subset       & $P(\sigma^n)$                   \\\hline
$\sigma^1$ & $p_0^0\rightarrow p_1^1\rightarrow p_2^1\rightarrow p_3^1\rightarrow d$ & $\{s_1\}$ & $0.563\times 0.836\times 0.881$ \\
$\sigma^2$ & $p_0^0\rightarrow p_1^0\rightarrow p_2^1\rightarrow p_3^1\rightarrow d$ &     $\{s_2\}$         & $0.437\times 0.652\times 0.881$ \\
$\sigma^3$ & $p_0^0\rightarrow p_1^0\rightarrow p_2^0\rightarrow p_3^1\rightarrow d$ &      $\{s_2\}$        & $0.437\times 0.348\times 1.000$ \\
$\sigma^4$ & $p_0^0\rightarrow p_1^1\rightarrow p_2^2\rightarrow p_3^2\rightarrow d$ &       $\{s_1,s_2\}$       & $0.563\times 0.164\times 1.000$ \\
$\sigma^5$ & $p_0^0\rightarrow p_1^1\rightarrow p_2^1\rightarrow p_3^2\rightarrow d$ &      $\{s_1,s_3\}$        & $0.563\times 0.836\times 0.119$ \\
$\sigma^6$ & $p_0^0\rightarrow p_1^0\rightarrow p_2^1\rightarrow p_3^2\rightarrow d$ &      $\{s_2,s_3\}$        & $0.437\times 0.652\times 0.119$ 
\end{tabular}
    \caption{Value function and path choice probabilities for the BiC DAG example.}
    \label{tab:bin-prob}
\end{table}
Similar to the BiC DAG, we illustrate the  MuC DAG representation of the toy instance in \autoref{fig:multi-example} where dashed arcs represent the termination of the choice process, i.e., the choice-maker finishes the selecting process and directly navigates to the destination. 
We report the value function as well as the path choice probabilities in \autoref{tab:multi-prob} which also shows that path choice probabilities are identical to the choice probabilities of the corresponding subsets.  

\begin{table}[htb]
    \centering
    \begin{tabular}{ll}
        \hline
        Node & $V^{\MuC}(\cdot)$    \\ \hline
        $p_0^0$     & -0.118 \\
        $p_1^1$     & 0.306 \\
        $p_2^1$     & 0.127 \\
        $p_3^1$     & 0 \\
        $p_2^2$     & 0 \\
        $p_3^2$     & 0 \\
        $d$         & 0 \\
    \end{tabular}
    \quad
   \begin{tabular}{l|l|l|l}
No.        & Path                  & Subset & $P(\sigma^n)$                     \\\hline
$\sigma^1$ & $p_0^0\rightarrow p_1^1\rightarrow d$                  &    $\{s_1\}$    & $0.563\times 0.736$               \\
$\sigma^2$ & $p_0^0\rightarrow p_2^1\rightarrow d$                  &   $\{s_2\}$     & $0.285\times 0.881$               \\
$\sigma^3$ & $p_0^0\rightarrow p_3^1\rightarrow d$                  &   $\{s_3\}$     & $0.152\times 1.000$               \\
$\sigma^4$ & $p_0^0\rightarrow p_1^1\rightarrow p_2^2\rightarrow d$ &   $\{s_1,s_2\}$     & $0.563\times 0.164\times 1.000$   \\
$\sigma^5$ & $p_0^0\rightarrow p_1^1\rightarrow p_3^2\rightarrow d$ &     $\{s_1,s_3\}$    & $0.563\times 0.100\times 1.000$  \\
$\sigma^6$ & $p_0^0\rightarrow p_2^1\rightarrow p_3^2\rightarrow d$ &     $\{s_2,s_3\}$    & $0.285\times 0.119\times 1.000$   
\end{tabular}
    \caption{Value function and path choice  probabilities for the MuC DAG representation.}
    \label{tab:multi-prob}
\end{table}

\subsection{Maximum Likelihood Estimation}
In this section, our focus is on estimating the RL model applied to the DAGs. We specifically delve into the calculation of the value functions $V^{\BiC}(\cdot)$ and $V^{\MuC}(\cdot)$, along with their gradients, as they play a crucial role in the maximum likelihood estimation \citep{MaiFrejinger22}. Our analyses primarily revolve around the BiC DAG, with a brief discussion for the MuC DAG case.

According to prior work \citep{FosgFrejKarl13,MaiFosFre15}, to compute the value function, it is convenient to use a vector $\bz$ of size $|\cN|$ with elements $z^{\BiC}_k = \exp\left(\frac{1}{\mu} V^{\BiC}(k)\right)$ for all $k\in \cN$. From Equation \ref{eq:value}, $\bz^{\BiC}$ satisfies the following system of  equations
\begin{equation}\label{eq:system-z-Bic}
z^{\BiC}_k =\begin{cases}
    1  &\text{ if } k =d \\
    \sum_{a\in N(k)} z^{\BiC}_a \exp\left(\frac{1}{\mu}v^{\BiC}(a|k)\right) & \text{ otherwise}. 
\end{cases} 
\end{equation}
By defining a matrix $\bM^{\BiC}$ of size $|\cN|\times|\cN|$ with entries
    \[
        M^{\BiC}_{ka} = \begin{cases}
           \exp{\left(\frac{1}{\mu} v^{\BiC}(a|k)\right)},    & a\in N(k), k\in \cN\backslash \{d\} \\
            0   & \text{otherwise}.
        \end{cases}
    \]
The linear system  in \eqref{eq:system-z-Bic} can be written in a matrix form as $\bz^{\BiC} = \bM^{\BiC}\bz^{\BiC} + \bb$ or $(\bI-\bM^{\BiC})\bz^{\BiC} = \bb$,  where $\bb$ is a vector of size $|\cN|$ with zero elements except $b_d = 1$ and $\bI$ is the identity matrix of size $|\cN|$. As a result, one can obtain $\bz^{\BiC}$ by solving the system of linear equations $(\bI-\bM^{\BiC})\bz^{\BiC} = \bb$ , which will always give a unique solution if $(\bI-\bM^{\BiC})$  is invertible. In a general setting where the graph may have cycles, it is possible that $(\bI-\bM^{\BiC})$ is not invertible if  the arc utilities $v^{\BiC}(a|k)$ are not negative and their the magnitudes are not significant \citep{MaiFrejinger22}. In the context of the BiC (or MuC) DAG,  the graph does not contain any cycles, implying that the matrix $(\bI-\bM^{\BiC})$ is consistently invertible, according to Proposition \ref{prop:I-M} below. Consequently, the vector $\bz^{\BiC}$ can be obtained by utilizing the inverse matrix of $(\bI-\bM^{\BiC})$, specifically as $\bz^{\BiC} = (\bI-\bM^{\BiC})^{-1}\bb$.
\begin{proposition}\label{prop:I-M}
    $(\bI-\bM^{\BiC})$ is always invertible. 
\end{proposition}
\proof{}
For any $n\in \mathbb{N}$, let us consider $(\bM^{\BiC})^n = \underbrace{\bM^{\BiC} \times \bM^{\BiC} \times \ldots \bM^{\BiC}}_\text{$n$ times}$ with entries
\[
(\bM^{\BiC})^n_{ss'} = \sum_{\substack{(t_0,\ldots,t_n) \\ t_i \in \cN,~ i = 0,\ldots
,n\\ t_0 = s, t_n = s'}} \left(\prod_{i=0}^{n-1}M^{\BiC}_{t_it_{i+1}}  \right).
\]
We see that the BiC DAG is cycle-free and any path in the DAG has no more than $m+2$ nodes. Thus, for any $n>m+1$,  for any sequence $(t_0,s_1,\ldots,t_n)$ there is at least a pair $(t_j, t_{j+1})$, $0\leq j\leq n-1 $ such that $t_{j+1}\notin N(t_j)$, or $M^{\BiC}_{t_it_{j+1}} = 0$, leading to the fact  that $(\bM^{\BiC})^n_{ss'}  = 0$ for any $s,s'\in \cN$. Thus, if $n>m+1$,  we have $(\bM^{\BiC})^n = \textbf{0}$ (i.e., the all-zero matrix of size $|\cN| \times |\cN|$). We then have
\[
 (\bI-\bM^{\BiC})\left(\sum_{t=0}^{n-1} (\bM^{\BiC})^t \right) = (\bI - (\bM^{\BiC})^n) =\bI,
\]
which implies $\det(\bI-\bM^\BiC) \neq 0$, or equivalently, $\bI-\bM^\BiC$ is invertible, as desired.  
\endproof

The invertibility of $\bI-\bM^\BiC$ implies that we can solve the linear system $(\bI-\bM^{\BiC})\bz^{\BiC} = \bb$ to obtain a unique solution. Directly inverting the matrix $(\bI-\bM^{\BiC})$ would be not practically viable if the size of $\bM^\BiC$ is large. An alternative method would be to use a simple value iteration, i.e., we iteratively compute  $(\bz^{\BiC,t+1}) = \bM^{\BiC}\bz^{\BiC,t} + \bb$ with $t=0,1,...$,  starting from an initial vector $\bz^{\BiC,0}$, until converging to a fixed point solution. In the context of the BiC DAG, it is possible to show that the value iteration will converge to the desired fixed-point solution after $m+2$ iterations from any starting vector $\bz^{\BiC,0}$.
\begin{proposition}\label{prop:convergence-bic}
For the BiC DAG, the value iteration solves the linear system $(\bI-\bM^{\BiC})\bz^{\BiC} = \bb$ after $m+2$ iterations. As a result, the value function can be computed in $\cO(m^2U)$. 
\end{proposition}
\proof{}
We leverage the fact that the BiC DAG is cycle-free and its nodes are organized into $m+2$ tiers and arcs only connect nodes in consecutive tiers. To facilitate the later arguments, let us denote $\cT(k)$, for any $k \in \cN$ as the index of the tier of node $k$, i.e., for any node $p^i_j\in \cN$, $\cT(p^i_j) = j$, and $\cT(d) = m+1$. Let $\bz^{\BiC*}$ be the  unique fixed point solution to the system $(\bI-\bM^{\BiC})\bz^{\BiC} = \bb$. We will prove, by induction, that after the $t$-th iteration of the value iteration, $z^{\BiC,t}_k = z^{\BiC*}_k$ for all $k\in\cN$ such that $\cT(k)\geq m+2-t$.  This, indeed, holds for $t=1$ as we see that  $z^{\BiC,t}_d = z^{\BiC*}_d$. Let's assume that it holds for $t\geq 1$, i.e., $z^{\BiC,t}_k = z^{\BiC*}_k$ for any $k$ such that $\cT(k)\geq m+2-t$. In the next iteration, we will update $\bz^{\BiC,t+1}$ as 
\[
z^{\BiC,t+1}_k = \sum_{a\in N(k)}z^{\BiC,t} M^{\BiC}_{ka} + b_k, ~\forall k\in \cN\backslash \{d\}. 
\]
we have that for any $k\in \cN$ such that $\cT(k)\geq m+2-(t+1)$ ($k$ in Tier $m+2-(t+1)$), then for any $a\in N(k)$ we have $\cT(a)\geq m+2-t$. From the induction assumption we have $z^{\BiC,t}_a = z^{\BiC*}_a$ for any $a\in N(k)$. Thus, for any $k$ such that $\cT(k)\geq m+2-(t+1)$,
\[
z^{\BiC,t+1}_k = \sum_{a\in N(k)}z^{\BiC,t} M^{\BiC}_{ka} + b_k = \sum_{a\in N(k)}z^{\BiC*} M^{\BiC}_{ka} + b_k = z^{\BiC*}_k,
\]
implying that the induction assumption holds for $t+1$. So after $m+2$ iterations, $z^{\BiC,t}_a = z^{\BiC*}_a$ for any $a\in \cN$.  

For the computational complexity, we see that $\bM^{\BiC}$ contains $|\cA| = \cO(mU)$ non-zero entries. Thus, each iteration $(\bz^{\BiC,t+1}) \leftarrow \bM^{\BiC}\bz^{\BiC,t} + \bb$ can be done in $\cO(mU)$. Consequently, we can obtain the fixed-point solution after $m+2$ iterations of the value iteration which can be done in $\cO(m^2U)$. 
\endproof

In the case of the MuC DAG, in a similar way, since the graph is cycle-free, we can prove that the matrix $\bI - \bM^\MuC$ is always invertible. The fixed-point solution can also be computed by 
value iteration as $(\bz^{\MuC,t+1}) \leftarrow \bM^{\MuC}\bz^{\BiC,t} + \bb$, $t = 0,1,...$. The following proposition states a result regarding the convergence of the value iteration and the computational complexity. 
\begin{proposition}\label{prop:convergence-muc}
For the MuC DAG, the value iteration solves the linear system $(\bI-\bM^{\MuC})\bz^{\MuC} = \bb$ after $m+2$ iterations, which can be done in  $\cO(m^3U)$. 
\end{proposition}
\proof{}
The claim can be validated in a similar way as the proof of Proposition \ref{prop:convergence-bic} as the nodes are also arranged into $m+2$ tiers. The only difference is that, in the MuC DAG, we have arcs connecting each node to several nodes in the subsequent tiers. This, however, does not affect the induction proof. That is, if after $t$ iterations we have $z^{\BiC,t}_k = z^{\BiC*}_k$ for any $k$ such that $\cT(k)\geq m+2-t$, then in the next iteration, we will update $\bz^{\BiC,t+1}_k$ using $\bz^{\BiC,t}_a$ such that $a\in N(k)$. The induction assumption also implies that  $\bz^{\BiC,t}_a = \bz^{\BiC*}_a$ for any $a\in N(k)$, leading to the result that the induction assumption also holds for $t+1$, which validates the convergence of the value iteration. For the computational complexity, we 
 see that $\bM^{\MuC}$ has $\cO(m^2U)$ non-zero elements. So, each iteration of the value function will take about $\cO(m^2U)$ time,  and in total, the value function will take $\cO(m^3U)$ time.  
\endproof

Once the linear system $(\bI-\bM^{\BiC})\bz^{\BiC} = \bb$ is resolved, the gradients of the value function w.r.t parameter $\beta_q$ can be computed as
  \[
    \frac{\partial\bz^\BiC}{\beta_q} = (\mathbf{I-M^\BiC})^{-1} \left[\frac{\partial\bM^\BiC}{\beta_q}\right] \bz^\BiC
\text{ and }   
    \frac{\partial V^{\BiC}(k)}{\partial \beta_q} = \mu \frac{1}{z^\BiC_k} \left(\frac{\partial z^\BiC_k}{\partial \beta_q}\right),
    \]
which is another system of linear equations.

\section{Nested Recursive Logit Model on the DAGs}\label{sec:NestedRL-model}
\cite{MaiFosFre15} propose the nested RL model to capture utility correlation in the network, overcoming the Independence of irrelevant alternatives (IIA) limitation of the RL model \citep{McFa81,FosgFrejKarl13}. One notable advantage of the nested RL approach is its ability to naturally capture correlation patterns that arise from the graph structure, similar to the network GEV model \citep{DalyBier06,MaiFreFosBas15_DynMEV}. This is achieved by allowing the scale parameters of the random terms to vary across nodes. In the context of multiple discrete choices, we have shown previously that there is a one-to-one mapping between a composite alternative to a path in the BiC or MuC graphs and the logit choice model over the subsets is equivalent to the RL model applied to either the BiC or MuC graphs. 
So, it is to be expected that the nested RL model on the BiC or MuC would be useful to naturally capture the correlation between composite alternatives' utilities. In this section, we introduced the nested RL applied to the DAGs and explore its properties through illustrative examples.


\subsection{Nested Recursive Logit}
We use the same notations as in the description of the RL model. In the context of the nested RL,  the expected maximum utility of a node $k\in \cN$ is also defined recursively  as
\begin{equation*}
    \widetilde{V}^{\BiC}(k)  = 
    \begin{cases}
    0,   & k = d \\
    -\infty, & N(k) = \emptyset\\
    \bbE \Bigg[\max_{a\in N(k)} \bigg\{v^\BiC(a|k) + \mu_k \epsilon(a) + \widetilde{V}^\BiC(a) \bigg\}\Bigg],   & \forall k\in\cN
    \end{cases}
\end{equation*}
The difference with regards to the RL model is that the scale parameters $\mu_k$ are now varying across nodes and can be modeled as functions of the node attributes.
Since the choice at each node is still the logit model, the value function can be computed through closed-form equations:
\begin{equation}\label{eq:nested-value}
    \exp{\left( \frac{1}{\mu_k} \widetilde{V}^\BiC(k) \right)} = 
    \begin{cases}
    1,   & k = d \\
    0,   & N(k) = \emptyset \\
    \sum_{a\in N(k)} \exp{\left( \frac{1}{\mu_k} (v^\BiC(a|k) + \widetilde{V}^\BiC(a)) \right)},   & \forall k\in \cN
    \end{cases}
\end{equation}
and the  node choice probabilities are computed as 
\[
\begin{aligned}
    P^\BiC(a|k) &= \frac{\exp{\left( \frac{1}{\mu_k} (v^\BiC(a|k) + \widetilde{V}^\BiC(a)) \right)}}{\sum_{a'\in N(k)} \exp{\left( \frac{1}{\mu_k} (v^\BiC(a'|k) + V^\BiC(a')) \right)}} \\
    &= \exp{\left( \frac{1}{\mu_k} (v^\BiC(a|k) + \widetilde{V}^\BiC(a) - \widetilde{V}^\BiC(k)) \right)}
\end{aligned}
\]
To compute the value function, we denote $\widetilde{\bz}^\BiC$ as a vector of size $|\cN|$ with entries $\widetilde{z}^\BiC_k = \frac{1}{\mu_k} \exp\left(\widetilde{V}^{\BiC}(k)\right)$ and rewrite \autoref{eq:nested-value} 
\begin{equation}\label{eq:nested-z}
\widetilde{z}^\BiC_k  = \sum_{a\in \cN} M^\BiC_{ka} (\widetilde{z}^\BiC_a)^{\mu_a/\mu_k} + b_k,~ \forall k\in \cN.     
\end{equation}
where matrix $\bM^\BiC (|\cN|\times |\cN|)$ is now defined as $\bM^\BiC_{ka} = \exp(v^\BiC(a|k)/\mu_k)$ for all $a\in N(k)$ and $\bM^\BiC_{ka}=0$ if $a\notin N(k)$, $\forall k\in \cN$, and $\bb$ is still a vector of size $|\cN|$ will zero elements except $b_d = 1$. The system of equations in \eqref{eq:nested-z} is highly nonlinear and is more intricate to solve compared to that of the RL model. \citep{MaiFosFre15} show that this system can be solved conveniently via value iteration for which the convergence cannot be guaranteed if the network contains cycles \citep{MaiFrejinger22}. In our context, it is possible to show that the value iteration always converges to a unique fixed point solution after $m+2$ iterations, for both  BiC and MuC graphs.   
\begin{proposition}
    For both BiC and MuC DAGs, the value iteration solves the system \eqref{eq:nested-z} after $m+2$ iterations with any starting vector $\widetilde{\bz}^{\BiC,0}$ (or $\widetilde{\bz}^{\MuC,0}$).
\end{proposition}
\proof{} 
The proof is very similar to the proof of Propositions \ref{prop:convergence-bic}  and \ref{prop:convergence-muc}. That is, since the nodes are organized in $(m+2)$ tiers, we can prove, by induction, that after $t$ iterations, the value iteration will update $\widetilde{\bz}^{\BiC,t}$ such that  $\widetilde{\bz}^{\BiC,t}_k =  \widetilde{\bz}^{\BiC*}_k$ for all $k\in \cN$ such that $\cT(k) = m+2-t$, where $\widetilde{z}^{\BiC*}$ the fixed point solution. As a result, the value iteration will return the fixed point solution after $m+2$ iterations. The same arguments apply to the MuC graph.  
\endproof

The gradients of the value function,  however, can be computed by solving systems of linear equations, similar to the case of the RL model. We refer the reader to \citep{MaiFosFre15} for more details.

\subsection{Example and Properties}
Within this section, we present examples to examine the characteristics of the nested RL model when it is applied to the DAGs. We, however, refrain from discussing how the nested RL model captures the correlation arising from the graph structure, as this has already been exploited in previous work \citep{MaiFosFre15}.
In the previous section, we show that for both BiC and MuC graphs,  the RL model is equivalent to the LMDC model, regardless of the order of the elemental alternatives that are chosen to construct the DAGs. In other words, the order of the alternatives does not affect the path choice probabilities given by the RL model. In the following, we will show that it might not be the case under the nested RL model, i.e. the path choice probabilities may change if we shuffle the order of the elemental alternatives when constructing the DAGs.  

\begin{figure}[htb]
    \centering    \includegraphics[width=1\linewidth]{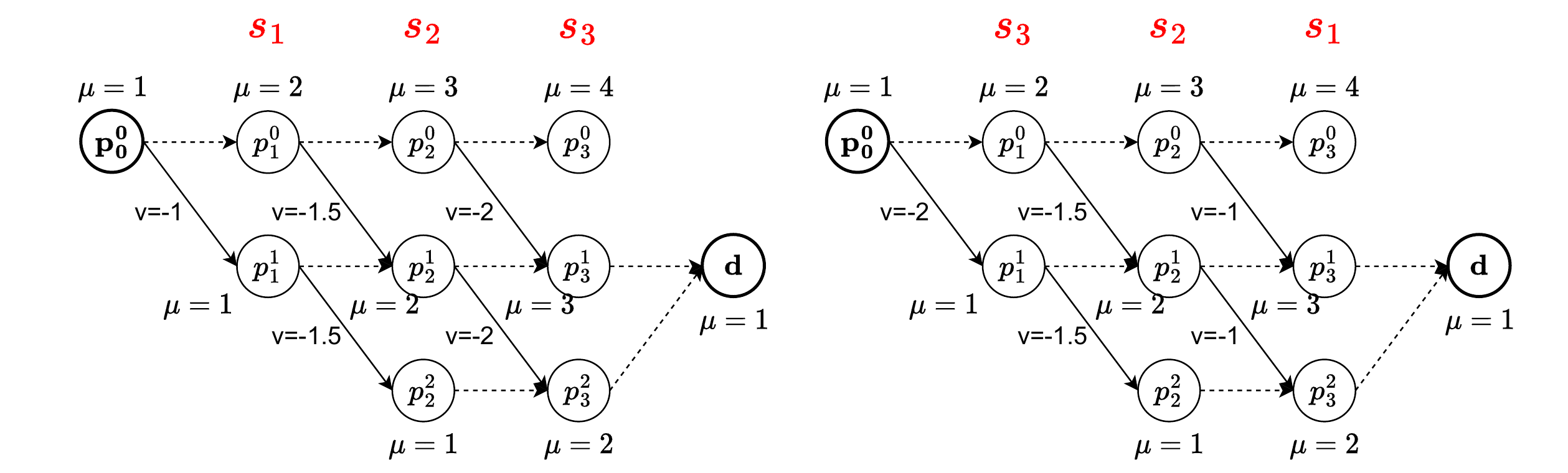}
    \caption{BiC DAG representations based on different alternative orders, for the 3-alternative choice problem.}
    \label{fig:bin-example-nested}
\end{figure}
We take the same example as in Section \ref{subsec:example}. There are 3 alternatives denoted as $\{s_1,s_2,s_3\}$ and each alternative $s$ is associated with an utility $v(s)$ and a scale parameter $\mu_s$. In \autoref{fig:bin-example-nested}
we illustrate two BiC DAG representations of the choices over the three alternatives, where the left-hand-side figure is constructed based on the order $\{s_1,s_2,s_3\}$
and the right-hand-side one is based on $\{s_3,s_2,s_1\}$.
We also provide the arc utilities and the scale parameters next to the corresponding arcs and nodes.  
To facilitate the later analysis, we denote the left-hand-side graph as BiC-1 and the other one as BiC-2.

We apply the nested RL model to the BiC-1 and BiC-2 graphs and report the path probabilities in \autoref{tab:nested-prob}, where each row reports a composite alternative, the mapping paths in BiC-1 and BiC-2 and their corresponding choice probabilities given by the nested RL. It can be observed that the nested RL offers different distributions over composite alternatives (or subsets) when applied to the BiC-1 and BiC-2 graphs, noting that the distributions are the same when using the RL model. 
This suggests that the chosen order of alternatives used to construct the BiC DAG representations would affect the resulting nested RL choice model.

\begin{table}[htb]
\centering
\begin{tabular}{l|l|l|l|l}
 & \multicolumn{2}{l|}{BiC-1} & \multicolumn{2}{l}{BiC-2} \\ 
\hline
Subset & Path & \multicolumn{1}{c|}{\begin{tabular}[c]{@{}c@{}}Path\\Probabilities\end{tabular}} & ~Path~ ~ & \multicolumn{1}{c}{\begin{tabular}[c]{@{}c@{}}Path\\probabilities\end{tabular}} \\ 
\hline
$\{s_1\}$ & $p_0^0\rightarrow p_1^1\rightarrow p_2^1\rightarrow p_3^1\rightarrow d$ & 0.28 & $p_0^0\rightarrow p_1^0\rightarrow p_2^0\rightarrow p_3^1\rightarrow d$ & 0.10 \\
$\{s_2\}$ & $p_0^0\rightarrow p_1^0\rightarrow p_2^1\rightarrow p_3^1\rightarrow d$ & 0.27 & $p_0^0\rightarrow p_1^0\rightarrow p_2^1\rightarrow p_3^1\rightarrow d$ & 0.29 \\
$\{s_3\}$ & $p_0^0\rightarrow p_1^0\rightarrow p_2^0\rightarrow p_3^1\rightarrow d$ & 0.21 & $p_0^0\rightarrow p_1^0\rightarrow p_2^0\rightarrow p_3^1\rightarrow d$ & 0.37 \\
$\{s_1,s_2\}$ & $p_0^0\rightarrow p_1^1\rightarrow p_2^2\rightarrow p_3^2\rightarrow d$ & 0.05 & $p_0^0\rightarrow p_1^0\rightarrow p_2^1\rightarrow p_3^2\rightarrow d$ & 0.01 \\
$\{s_1,s_3\}$ & $p_0^0\rightarrow p_1^1\rightarrow p_2^1\rightarrow p_3^2\rightarrow d$ & 0.10 & $p_0^0\rightarrow p_1^1\rightarrow p_2^1\rightarrow p_3^2\rightarrow d$ & 0.06 \\
$\{s_2,s_3\}$ & $p_0^0\rightarrow p_1^0\rightarrow p_2^1\rightarrow p_3^2\rightarrow d$ & 0.10 & $p_0^0\rightarrow p_1^0\rightarrow p_2^1\rightarrow p_3^2\rightarrow d$ & 0.17\\\hline
\end{tabular}
  \caption{Path choice probabilities given by the nested RL on the BiC-1 and BiC-2 graphs.}
    \label{tab:nested-prob}
\end{table}

\begin{figure}[htb]
    \centering    \includegraphics[width=0.6\linewidth]{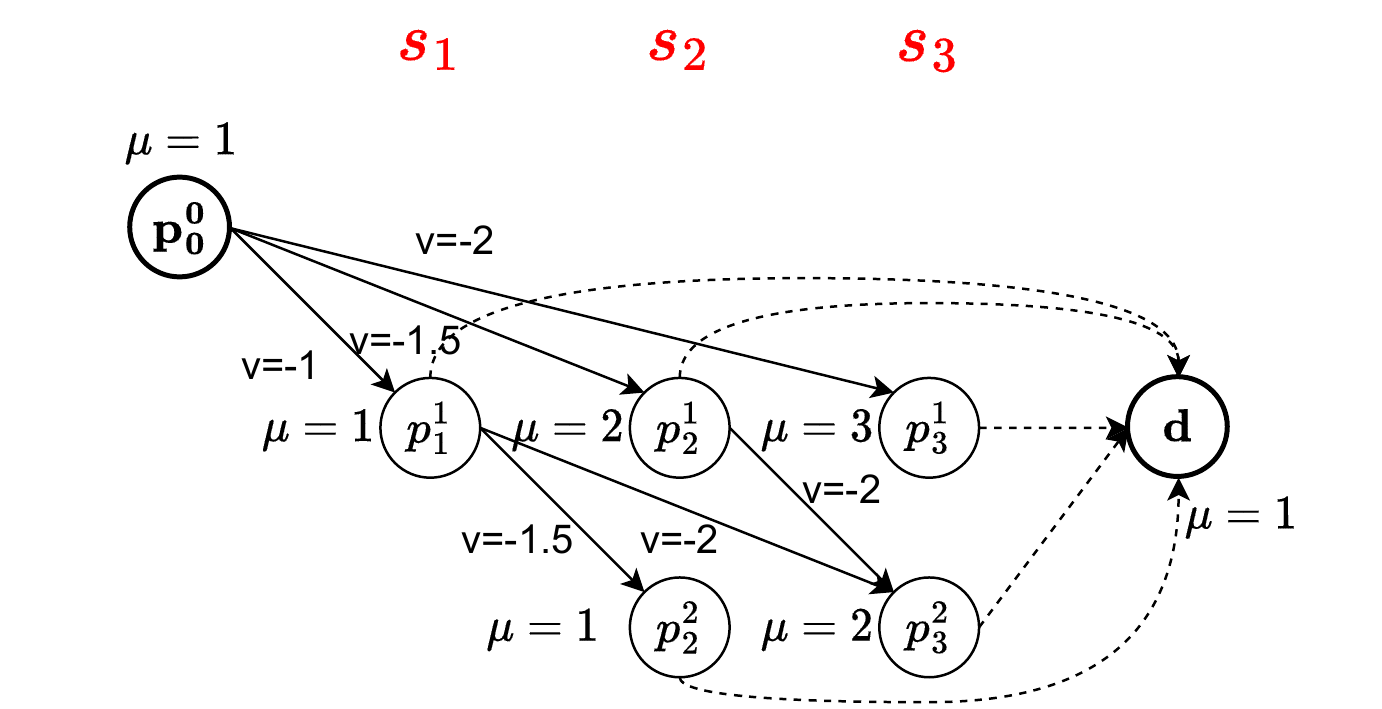}
    \caption{MuC DAG representation for the 3-alternative choice problem.}
    \label{fig:multi-example-nested}
\end{figure}

\begin{table}[htb]
\centering
\begin{tabular}{l|l|l}
Subset & Path & \multicolumn{1}{c}{\begin{tabular}[c]{@{}c@{}}Path\\Probabilities\end{tabular}} \\ 
\hline
$\{s_1\}$  & $p_0^0\rightarrow p_1^1\rightarrow d$      & 0.35 \\
$\{s_2\}$  & $p_0^0\rightarrow p_2^1\rightarrow d$      & 0.29 \\
$\{s_3\}$  & $p_0^0\rightarrow p_3^1\rightarrow d$      & 0.13 \\
$\{s_1, s_2\}$  & $p_0^0\rightarrow p_1^1\rightarrow p_2^2\rightarrow d$      & 0.08 \\
$\{s_1, s_3\}$  & $p_0^0\rightarrow p_1^1\rightarrow p_3^2\rightarrow d$      & 0.05 \\
$\{s_2, s_3\}$  & $p_0^0\rightarrow p_2^1\rightarrow p_3^2\rightarrow d$      & 0.11 \\ \hline
\end{tabular}
  \caption{Path choice probabilities given by the nested RL on the MuC graph.}
    \label{tab:nested-prob2}
\end{table}

Using the same example, we can also show that the MuC DAG with different alternative orders can yield different nested RL models. It is more interesting to explore the question that if the BiC and MuC are constructed using the same alternative order, does the probability distribution over paths remains the same? To this end, we simply keep the same order $\{s_1,s_2,s_3\}$ to construct the MuC graph (as illustrated in \autoref{fig:multi-example-nested}). The path probabilities are reported in   
 \autoref{tab:nested-prob2}, which clearly shows that the path distribution given by the nested RL on the MuC DAD is different from those given by the nested RL and BiC reported in  \autoref{tab:nested-prob}. So,  we conclude that, in general,  the probability distributions given by the BiC DAG and MuC DAG models, based on the same alternative order, are not necessarily the same. Interestingly, it can be shown that, under some specific settings of the scale function, these two distributions can be completely identical. We state this result in Proposition \ref{prop:nested-util} below.

\begin{proposition}\label{prop:nested-util}
    Assume that the  BiC DAG and  MuC DAG are constructed based on the same order of elemental alternatives and the scale parameter $\mu_{p_j^c}$ associated with node $p_j^c$ in both BiC and MuC graphs is only dependent on $c$ (i.e., the number of selected alternatives at that node). Then the nested RL models on the  BiC and MuC are equivalent, i.e., for any subset $S\in\Omega^{[L,U]}$ we have
    \[
    \widetilde{P}^{\BiC}(\sigma^{\BiC}(S)) =   \widetilde{P}^{\MuC}(\sigma^{\MuC}(S)).
    \]
\end{proposition}
The proof can be found in the appendix. 

    

\section{Numerical Experiments}\label{sec:exp}
To evaluate the performance of the two proposed DAGs for the multiple discrete choice problem, we compare them against two logit-based baseline models. 
These baseline models will be described in the subsequent sections.
In this comparison, we focus solely on evaluating our models and approaches against logit-based models. This choice is made because the other existing choice models developed for multiple choice analysis do not align with the RUM and bounded rationality principles.  
 The experiments are conducted using both synthetic instances and two real-world datasets. In the following subsections, we will begin by describing the baseline models. Subsequently, we will present the numerical results conducted with generated instances under various settings. Finally, we will show our experiments conducted using two real-world datasets. All experiments are performed on a Google Cloud TPU v2-8 server with a 96-core Intel Xeon 2.00 GHz CPU. 


\subsection{Baseline models}
In the following, we present some simple logit-based baseline models that can be used for the multiple discrete choice problem.

\textbf{{Single-Choice model (SC-Base). }}
The first baseline model is formulated by considering each composite alternative as a  sequence of multiple independent single choices. As a result, the choice probability of a composite alternative can be computed as the product of single-choice probabilities. Specifically, the choice probability of a composite alternative $S$  can be computed as
\[
P(S) = \prod_{i\in S} P(i|[m]) = \prod_{i\in S} \frac{e^{\frac{1}{\mu}v_i}}{\sum_{i'\in [m]} e^{\frac{1}{\mu}v_{i'}}}
\]
It can be seen that the SC-Base model always gives lower choice probabilities, compared to the LMDC model. Furthermore, the SC-Base model is unable to incorporate the lower and upper constraints on the number of selected elemental alternatives, primarily due to its simple structure.


\textbf{{Multi-Choice with sampled choice sets (MC-Base).}} As mentioned earlier, a challenge raised from the use of the LMDC model is the impracticality of exhaustively enumerating the composite alternatives within the choice set $\Omega^{[L,U]}$. A simple way to deal with this challenge is to replace $\Omega^{[L,U]}$ by a small set of sampled alternatives. This modeling approach would share some similarities with path-based models in route choice modeling -- since there are exponentially many paths connecting an origin and a destination, one can sample paths to make the estimation practical \citep{Prat09}. One limitation of such a sampling-based approach is that the estimation may be not consistent, as it relies on the specific choice sets that are sampled.     

Under the MC-Base model, we can sample composite alternatives from $\Omega^{[L,U]}$ that satisfy the bound constraints (i.e., the size 
 is within $[L,U]$). We then use this sampled choice set to compute the choice probabilities, and  to estimate the choice parameters. Specifically, let $\Gamma$ be a sampled choice set of feasible composite alternatives. We compute the choice probability of a composite alternative $S\subset [m]$ as   
\[
    P(S|\bbt) = \frac{\exp \left( \frac{1}{\mu} \sum_{i\in S} v_i(\bbt) \right)}{\sum_{S'\in \Gamma} \exp \left( \frac{1}{\mu} \sum_{i\in S'} v_i(\bbt) \right)}. \\
\]
It can be seen that the MC-Base model always gives a higher choice probability for each composite alternative, compared to that from the LMDC model, as the sampled choice set $\Gamma$ is just a subset of $\Omega^{[L,U]}$. Consequently, if there is an alternative $S$ that is not included in $\Gamma$, the MC-Base will assign it a zero probability. This would be an issue if $S$ is an actual choice made by the choice-makers.

\subsection{Synthetic Data}
The objective here is to evaluate the losses incurred by employing the two baseline models, SC-Base and MC-Base, when the true model is the LMDC model. Additionally, we aim to measure the estimation times required by different approaches.
To this end, we generate various problem instances by taking $m\in\{5, 10, 15, 20, 30, 50\}$  and vary the upper bound $U$ as $U=3$ for $m=5$, $U\in\{3, 5\}$ for $m=10$, $U\in\{5, 10\}$ for $m=15$, $U\in\{5, 10\}$ for $m=20$, $U\in\{10, 20\}$ for $m=30$, and  $U\in\{10, 30\}$ for $m=50$. We also set $L=0$ for all the instances. 
For each choice of $m$, we randomly generate three attributes for each elemental alternative. We also choose a vector $\bbt^0$  and use it to define a ``ground-truth'' model to generate observations. For each choice model (a baseline or the LMDC model), we generate a set of observations for the estimation and use the parameter estimates to compute the log-likelihood of another set of observations. By comparing these likelihood values, we aim to assess the out-of-sample performances of the considered models. For the MC-Base model, the estimation requires sampled choice sets of composite alternatives. To have this, we simply collect subsets from the observations. In Table \ref{tab:size-MC-Base} below we report the average sizes of the generated choice sets used for estimating the MC-Base model.
 
\begin{table}
\centering
\caption{Average sizes of the sampled choice sets used  for estimating the MC-Base model.}
\label{tab:size-MC-Base}
\begin{tabular}{c|c|c|cc|cc|cc|cc}
\multirow{2}{*}{Dataset} & $m$ & 5 & \multicolumn{2}{c|}{10} & \multicolumn{2}{c|}{20} & \multicolumn{2}{c|}{30} & \multicolumn{2}{c}{50} \\ 
\cline{2-11}
 & $[L,U]$ & {[}0,3] & {[}0,3] & {[}0,5] & {[}0,5] & {[}0,10] & {[}0,10] & {[}0,20] & {[}0,10] & {[}0,30] \\ 
\cline{1-11}
\multicolumn{2}{c|}{\begin{tabular}[c]{@{}c@{}}Size of sampled \\choice set\end{tabular}} & \multicolumn{1}{l|}{26} & \multicolumn{1}{l}{176} & \multicolumn{1}{l|}{630} & \multicolumn{1}{l}{2799} & \multicolumn{1}{l|}{2993} & \multicolumn{1}{l}{3000} & \multicolumn{1}{l|}{3000} & \multicolumn{1}{l}{3000} & \multicolumn{1}{l}{3000} \\
\hline
\end{tabular}
\end{table}



\begin{table}[htb]
    \centering
        \caption{Parameters used for generating synthetic data.}
    \label{tab:params}
    \begin{tabular}{l|r}
        \hline
        \multicolumn{1}{c|}{\textbf{Parameter}}   & \multicolumn{1}{c}{\textbf{Value}} \\ \hline
        Number of attributes    & 3 \\
        $1^{st}$ attribute      & log-normal, mean=0, scale=1 \\
        $2^{nd}$ attribute      & log-normal, mean=0, scale=1 \\
        $3^{rd}$ attribute      & constant, value=1 \\
        $\bbt^0$   & $[-0.5, -0.02, -0.1]$ \\
        Size of the estimation set for SC-Base and MC-Base & 3000 \\
        Size of the estimation set for RL models & 1000 \\
        Size of the prediction set   & 300 \\ \hline 
    \end{tabular}
\end{table}

 Table \ref{tab:params} gives the parameters used for generating the synthetic instances and provides details regarding the sizes of the estimation and prediction sets. Each problem instance is randomly generated, consisting of three attribute values. The first two attributes follow a log-normal distribution with a mean of 0 and a scale of 1. The third attribute remains constant with a value of one. This attribute aims to reflect a natural phenomenon where the likelihood of people choosing a subset decreases as the number of items in that subset increases. In real-world datasets, it has been observed that a significant proportion of customers only purchase 1-2 products at a time.


We generate the synthetic instances as follows.
For each group, $(m, L,U)$, 300 sets of sample utilities are randomly generated. Then, for each set of alternative utilities, we generate 40 sets of observations using the true model with $\bbt^0$. Each choice model is then estimated using the generated sets of observations. Each vector of parameter estimates $\widehat{\bbt}$  obtained is then used by the true model to calculate the predicted log-likelihood value of the prediction set. The average predicted log-likelihood values and standard errors over 40 runs are reported in Table \ref{tab:synth} where the numbers before ``$\pm$'' are the means and those after ``$\pm$'' are the standard errors. The last column reports the average log-likelihood values given by the true model (i.e., the LMDC model) with the true parameters $\bbt^0$. 
The results clearly show that the average predicted log-likelihood values given by the RL model applied to the BiC and MuC graphs are almost identical to those returned by the true model, which numerically confirms the equivalence between the LMDC and RL model on the DAGs,  shown is Section~\ref{subsec:model-equal}. The SC-Base model performs the worst among the 4 models considered, especially when $m$ and $U$ are large. On the other hand, 
the predicted log-likelihood values obtained from the MC-Base model exhibit a high degree of accuracy, particularly for smaller instances (e.g. $m\leq 10$). However, as the value of $m$ increases, the discrepancies between the predicted log-likelihood values and the true values become more significant. Intuitively, when $m$ is small, there is a higher likelihood that the sampled choice set $\Gamma$ used in the MC-Base model closely resembles the universal choice set $\Omega^{[L,U]}$. However, as $m$ grows larger, the gap between the two sets becomes more pronounced.


\begin{table}[htb]
    \centering
        \caption{Predicted average log-likelihood values for synthetic data.}
    \label{tab:synth}
\begin{tabular}{cc|rrrr|r}
\multicolumn{2}{c|}{Dataset}                       & \multicolumn{4}{c|}{Model}                                                                                           & \multicolumn{1}{c}{\multirow{2}{*}{True model}} \\ \cline{1-6}
\multicolumn{1}{c|}{$m$}                 & $[L,U]$ & \multicolumn{1}{c}{RL+BiC} & \multicolumn{1}{c}{RL+MuC} & \multicolumn{1}{c}{MC-Base} & \multicolumn{1}{c|}{SC-Base} & \multicolumn{1}{c}{}                            \\ \hline
\multicolumn{1}{c|}{5}                   & [0,3]   & $-2.95\pm 0.04$            & $-2.96\pm 0.04$            & $-2.95\pm 0.02$             & $-6.30\pm 0.12$              & -2.09                                           \\ \hline
\multicolumn{1}{c|}{\multirow{2}{*}{10}} & [0,3]   & $-4.71\pm 0.06$            & $-4.70\pm 0.06$            & $-4.70\pm 0.03$             & $-9.87\pm 0.27$              & -3.18                                           \\
\multicolumn{1}{c|}{}                    & [0,5]   & $-5.98\pm 0.06$            & $-5.99\pm 0.06$            & $-6.01\pm 0.03$             & $-19.85\pm 0.76$             & -4.40                                           \\ \hline
\multicolumn{1}{c|}{\multirow{2}{*}{20}} & [0,5]   & $-9.40\pm 0.07$            & $-9.39\pm 0.07$            & $-9.94\pm 0.02$             & $-30.59\pm 4.68$             & -7.32                                           \\
\multicolumn{1}{c|}{}                    & [0,10]  & $-11.94\pm 0.08$           & $-11.97\pm 0.11$           & $-13.33\pm 0.00$            & $-97.05\pm 11.98$            & -7.36                                           \\ \hline
\multicolumn{1}{c|}{\multirow{2}{*}{30}} & [0,10]  & $-16.92\pm 0.09$           & $-16.92\pm 0.09$           & $-17.79\pm 0.00$            & $-60.54\pm 4.62$             & -13.71                                          \\
\multicolumn{1}{c|}{}                    & [0,20]  & $-19.05\pm 0.10$           & $-19.04\pm 0.09$           & $-20.77\pm 0.00$            & $-189.22\pm 23.31$           & -13.80                                          \\ \hline
\multicolumn{1}{c|}{\multirow{2}{*}{50}} & [0,10]  & $-22.63\pm 0.08$           & $-22.66\pm 0.09$           & $-23.32\pm 0.00$            & $-44.61\pm 2.63$             & -19.98                                          \\
\multicolumn{1}{c|}{}                    & [0,30]  & $-32.17\pm 0.11$           & $-32.18\pm 0.14$           & $-34.60\pm 0.00$            & $-243.53\pm 43.35$           & -23.86                                          \\ 
\end{tabular}
\end{table}

As claimed earlier, 
one of the main advantages of our approach is that the LMDC model becomes tractable to estimate via dynamic programming. We numerically assess this advantage by reporting the average estimation times required by the four models in \autoref{tab:synth-time}.  The SC-Base model is the fastest, as expected, due to its simple structure. The CPU times required by the MC-Base model strongly depend on the size of the sampled choice set and are quite unstable. Moreover, the estimation times for the two RL models on the BiC and MuC graphs vary compared to the MC-Base model. In certain cases, the RL models exhibit slower estimation times, while in other cases, they are faster. Nevertheless, the overall estimation times of the RL model are surprisingly short, even for larger instances with $m=50$. For instance, in cases where $m=50$ and $[L,U]=[0,30]$, it takes approximately 1.5 seconds to estimate the RL with the BiC DAG and around 4 seconds with the MuC DAG.
Additionally, we observe that the estimation of the RL model on the BiC DAG is faster compared to the MuC DAG. This discrepancy can be attributed to the fact that, although the number of nodes in both graphs is similar, the MuC graph contains more arcs and is denser in structure.



\begin{table}[htb]
    \centering
        \caption{Estimation time (in seconds) for synthetic data.}
    \label{tab:synth-time}
\begin{tabular}{cc|rrrr}
\multicolumn{2}{c|}{Dataset}                       & \multicolumn{4}{c}{Model}                                                                                           \\ \hline
\multicolumn{1}{c|}{$m$}                 & $[L,U]$ & \multicolumn{1}{c}{RL+BiC} & \multicolumn{1}{c}{RL+MuC} & \multicolumn{1}{c}{MC-Base} & \multicolumn{1}{c}{SC-Base} \\ \hline
\multicolumn{1}{c|}{5}                   & [0,3]   & $0.10\pm 0.03$             & $0.09\pm 0.02$             & $0.52\pm 0.12$              & $0.01\pm 0.00$              \\ \hline
\multicolumn{1}{c|}{\multirow{2}{*}{10}} & [0,3]   & $0.14\pm 0.01$             & $0.08\pm 0.01$             & $0.65\pm 0.10$              & $0.02\pm 0.00$              \\
\multicolumn{1}{c|}{}                    & [0,5]   & $0.14\pm 0.01$             & $0.10\pm 0.01$             & $0.97\pm 0.10$              & $0.04\pm 0.00$              \\ \hline
\multicolumn{1}{c|}{\multirow{2}{*}{20}} & [0,5]   & $0.24\pm 0.02$             & $0.19\pm 0.02$             & $1.79\pm 0.17$              & $0.07\pm 0.01$              \\
\multicolumn{1}{c|}{}                    & [0,10]  & $0.26\pm 0.02$             & $0.30\pm 0.03$             & $2.17\pm 0.27$              & $0.11\pm 0.01$              \\ \hline
\multicolumn{1}{c|}{\multirow{2}{*}{30}} & [0,10]  & $0.40\pm 0.01$             & $0.57\pm 0.03$             & $0.56\pm 0.30$              & $0.12\pm 0.01$              \\
\multicolumn{1}{c|}{}                    & [0,20]  & $0.61\pm 0.02$             & $0.99\pm 0.07$             & $0.69\pm 0.02$              & $0.18\pm 0.02$              \\ \hline
\multicolumn{1}{c|}{\multirow{2}{*}{50}} & [0,10]  & $0.67\pm 0.04$             & $1.58\pm 0.12$             & $0.51\pm 0.01$              & $0.13\pm 0.01$              \\
\multicolumn{1}{c|}{}                    & [0,30]  & $1.50\pm 0.81$             & $3.97\pm 0.30$             & $1.10\pm 0.03$              & $0.29\pm 0.04$              \\ 
\end{tabular}
\end{table}

\subsection{Real-world Datasets}
In this section, we compare the performances of different logit-based models based on two datasets extracted from real-world data sources. We also consider the two logit-based baselines -- the SC-Base and MC-Base models. In-sample results (i.e., estimation results) and out-of-sample results will be reported for each dataset. In the out-of-sample evaluation, the observations are randomly split into two sets with a predetermined ratio. Specifically, 80\% of the observations are utilized for parameter estimation, while the remaining 20\% serve as a holdout set to assess the predictive performance of the estimated parameters. We generate 40 estimation-holdout pairs and calculate the average log-likelihood values for each model, based on the universal choice set of composite alternatives (i.e.,  the LMDC model), which will be reported as the evaluation metric.



\textbf{{Jewelry-Store dataset.}}
This dataset\footnote{\url{https://www.kaggle.com/datasets/mkechinov/ecommerce-purchase-history-from-jewelry-store}} contains purchase data from December 2018 to December 2021 (3 years) from a medium-sized jewelry online store, collected by Open CDP project \citep{kaggle2021jewelry}. Each jewelry product has three attributes: \textit{Quantity} (total number of that product sold), \textit{Price, Popularity} (number of unique customers who purchased that product), and \textit{Gender} (which gender for which the product is designed).
We also include a constant attribute which takes values of 1 for all the products. This attribute will capture the number of selected products in each composite observation. Moreover, in order to enhance the accuracy of the simulation and prioritize significant products, we narrow our focus to data collected during the initial two years. Furthermore, we only consider products that have sold a minimum of 50 units.  In total, there are 436 items, and the dataset consists of 34,291 observations.
It is worth noting that approximately 93\% of the orders comprise no more than 5 products, while any orders exceeding this threshold are treated as anomalous ones.
As the dataset does not provide information regarding the lower and upper bounds on the number of selected items that customers consider when making choices, we attempt to infer these bounds based on the observations. That is, we partition the observations into three categories that correspond to customers purchasing 1-2 items, 3-5 items, and more than 5 items. This yields three sets of $(L,U)$ values, which are used to construct the DAGs. For the LMDC model, we utilize the BiC DAG, noting that the MuC DAG will yield the same estimation results.
For the MC-Base baseline, we collect subsets from the observations to form the choice set $\Gamma$, giving us a consideration set consisting of 3487 composite alternatives. 
The estimation outcomes are presented in Table \ref{tab:jewelry-estimate}. From the results, we can see that the parameter estimates for \textit{Quantity} and \textit{Price} seem to have expected signs. Moreover, parameter estimates for \textit{Quantity} and \textit{Const} are highly significant, while the estimates for \textit{Price} are not. This could be because the products have similar prices to each other, leading customers to prioritize the total quantity of items they can buy based on their budgets.
Based on intuition, as the \textit{Const} attribute represents the number of purchased items, it is expected that its estimates would have negative signs. This remark suggests that the parameter estimates of \textit{Const} provided by MC-Base and LMDC have the correct signs, whereas this is not the case for SC-Base. Finally, it is important to highlight that the log-likelihood values, obtained by utilizing the corresponding parameter estimates with the universal choice set $\Omega^{[L,U]}$, are presented in the final column of \autoref{tab:jewelry-estimate}. These values serve as an indication that the LMDC model significantly outperforms other models in terms of in-sample fit. Here,  it is important to note that the final log-likelihood values obtained through Maximum Likelihood Estimation (MLE) are not reported since they are computed based on different choice sets and therefore cannot be directly compared. Instead, to ensure a meaningful in-sample fit comparison, we employ the parameter estimates to calculate the log-likelihood values using the same choice set of composite alternatives, which is the universal choice set. 




\begin{table}[htb]
    \centering
    \caption{Estimation results for the Jewelry-Store dataset.}
    \label{tab:jewelry-estimate}
    \begin{tabular}{cl|rrrr|r}
    \multicolumn{2}{c|}{\multirow{2}{*}{Model}} & \multicolumn{4}{c|}{Attribute}                   & \multicolumn{1}{c}{\multirow{2}{*}{$\cL(\widehat{\bbt})$}} \\ \cline{3-6}
    \multicolumn{2}{c|}{} & \multicolumn{1}{c}{Quantity} & \multicolumn{1}{c}{Price} & \multicolumn{1}{c}{Gender} & \multicolumn{1}{c|}{Const} & \multicolumn{1}{c}{}       \\ \hline
    \multirow{3}{*}{SC-Base}     & $\widehat{\bbt}$      & 2.352  & -0.143                    & 0.005& 9.136& \multirow{3}{*}{-4,967,533}   \\
& Std. Err.    & 0.021  & 0.045                     & 0.016& 8.075&                 \\
& $t$-test(0)  & 111.925& -3.181                    & 0.341& 1.131&                 \\ \hline
    \multirow{3}{*}{MC-Base}     & $\widehat{\bbt}$      & 1.667  & -0.181                    & -0.059                     & -3.994                     & \multirow{3}{*}{-503,396}    \\
& Std. Err.    & 0.022  & 0.062                     & 0.016& 0.022&                 \\
& $t$-test(0)  & 74.263 & -2.897                    & -3.686                     & -181.578                   &                 \\ \hline
    \multirow{3}{*}{LMDC}   & $\widehat{\bbt}$      & 2.363  & -0.144                    & 0.005& -6.527                     & \multirow{3}{*}{-270,061}    \\
& Std. Err.    & 0.024  & 0.062                     & 0.018& 0.019&                 \\
& $t$-test(0)  & 98.704 & -2.312                    & 0.280& -343.973                   &                 \\ 
    \end{tabular}
\end{table}

Table \ref{tab:prediction-jewelry} reports the out-of-sample evaluation for the Jewelry-Store dataset. Similar to the synthetic data, we split the observation set into two parts, one part taking 80\% of the observations is utilized for parameter estimation, while the second part taking 20\% serves as a holdout set to assess the prediction performance of the estimated parameters. We estimate each model considered (LMDC, SC-Base or MC-Base) using the estimation set,  and employ the obtained parameters to calculate the log-likelihood value of the holdout set based on the universal choice set of composite alternatives $\Omega^{[L,U]}$. Because the data does not provide information about the lower and upper limits of customers' purchasing items, and these limits may vary for different customers, we categorize the observations based on the actual number of products purchased. For example, in Table \ref{tab:prediction-jewelry}, the first row presents the prediction results solely based on observations of a single selected item, and the row indicating $[L,U] = [1,11]$ reports results for observations where the number of purchased products ranges from 1 to 11. We indicate in bold the best log-likelihood value in each row. 

It can be seen, from Table \ref{tab:prediction-jewelry}, that when $L=U$, i.e., the number of purchased items is fixed, the predicted log-likelihood values are similar across the three models. When we increase the range $[L,U]$, the LMDC model begins to outperform other models, and the SC-Base becomes the worst. Overall, we see that the LMDC model provides significantly better prediction results compared to the two baselines. 

\begin{table}[htb]
\centering
\caption{Out-of-sample results for the Jewelry-Store dataset}
\label{tab:prediction-jewelry}
\begin{tabular}{c|c|crr}
$[L,U]$ & \begin{tabular}[c]{@{}c@{}}Number of  \\Observations\end{tabular} & LMDC & \multicolumn{1}{c}{MC-Base} & \multicolumn{1}{c}{SC-Base} \\ 
\hline
{[}1,1] & 30286 & \textbf{-5.97} & \textbf{-5.97} & \textbf{-5.97} \\
{[}2,2] & 3150 & \textbf{-11.22} & -11.41 & \textbf{-11.22} \\
{[}3,3] & 615 & -\textbf{16.18} & -16.45 & \textbf{-16.18} \\
{[}4,4] & 150 & \textbf{-20.98} & -21.12 & \textbf{-20.98} \\
{[}5,5] & 56 & \textbf{-24.92} & -25.58 & \textbf{-24.92} \\
{[}1,2] & 33436 & \multicolumn{1}{r}{\textbf{-6.78}} & -7.91 & -19.85 \\
{[}3,5] & 821 & \multicolumn{1}{r}{\textbf{-18.33}} & -25.56 & -43.85 \\
{[}6,10] & 33 & \multicolumn{1}{r}{\textbf{-31.79}} & -45.59 & -87.09 \\
{[}1,11] & 34291 & \multicolumn{1}{r}{\textbf{-7.24}} & -16.03 & -149.24
\end{tabular}
\end{table}


\textbf{{Book-Crossing dataset.}}
The dataset\footnote{\url{http://www2.informatik.uni-freiburg.de/~cziegler/BX/}} were crawled in 4 weeks by Cai-Nicolas Ziegler and coauthors from BookCrossing\footnote{\url{http://www.bookcrossing.com}}, surveying books purchased and customer ratings from Amazon Web Services. It is used as the benchmark data in the publication by the same authors \citep{ziegler2005improving}. Each book has three attributes: \textit{overall rating} from customers (only count explicit ones), \textit{popularity} (total number of customer ratings, both explicit and implicit), and its \textit{lifespan} (from the first time it is published to the date it was crawled).
To capture the number of selected items, we also include an additional attribute that is equal to 1 for every book. We name this attribute as ``\textit{Const}''.
The observations don't include some outlier samples where customers were unidentified or had bugged data. There are 100 books, the data consists of 16,989 observations, and the number of selected items varies from 1 to 70.  For the MC-Base baseline, we gather subsets from the observations to construct the choice set $\Gamma$. This results in a consideration set comprising 4886 composite alternatives.

\begin{table}[htb]
    \centering
        \caption{Estimation results for the Book-Crossing dataset}
    \label{tab:book-estimate}
    \begin{tabular}{cl|rrrr|c}
    \multicolumn{2}{c|}{\multirow{2}{*}{Model}} & \multicolumn{4}{c|}{Attribute}                   & \multicolumn{1}{c}{\multirow{2}{*}{$\cL(\widehat{\bbt})$}} \\ \cline{3-6}
    \multicolumn{2}{c|}{} & \multicolumn{1}{c}{Rating} & \multicolumn{1}{c}{Popularity} & \multicolumn{1}{c}{Lifespan} & \multicolumn{1}{c|}{Const} & \multicolumn{1}{c}{}       \\ \hline
    \multirow{3}{*}{SC-Base}     & $\widehat{\bbt}$      & 0.791  & 2.677                     & -0.079                     & 11.496                     & \multirow{3}{*}{-}  \\
& Std. Err.    & 0.030  & 0.028                     & 0.019& 12.085                     &                 \\
& $t$-test(0)  & 26.502 & 94.617                    & -4.095                     & 0.951&                 \\ \hline
    \multirow{3}{*}{MC-Base}     & $\widehat{\bbt}$      & -0.145 & 0.517                     & 0.115& -0.533                     & \multirow{3}{*}{-789,434}    \\
& Std. Err.    & 0.037  & 0.033                     & 0.018& 0.032&                 \\
& $t$-test(0)  & -3.904 & 15.706                    & 6.392& -16.775                    &                 \\ \hline
    \multirow{3}{*}{LMDC}   & $\widehat{\bbt}$      & 0.777  & 2.826                     & -0.078                     & -4.464                     & \multirow{3}{*}{-195,208}    \\
& Std. Err.    & 0.047  & 0.040                     & 0.022& 0.038&                 \\
& $t$-test(0)  & 16.590 & 71.192                    & -3.625                     & -118.189                   &                 \\ 
    \end{tabular}
\end{table}

The estimation results are reported in \autoref{tab:book-estimate} which show that most of the parameter estimates are significantly different from zero, except the estimate of \textit{Const} of the SC-Base. It is reasonable to anticipate that the sign of the \textit{Rating} parameter should be negative, as higher ratings should positively affect the attractiveness of a book. Consequently, the parameter estimates for \textit{Rating} exhibit correct signs in the LMDC and SC-Base models, while they display incorrect signs in the MC-Base model. The parameter estimates for \textit{Popularity} are, as expected, positive for all the models. The estimates for the attribute \textit{Const} provided by the LMDC model are negative and highly significant. In contrast, the estimate given by the MC-Base model is less significant, and the estimate given by the SC-Base model is even positive, which contradicts our prior intuition regarding the effect of the number of selected items on the item utilities.  The log-likelihood values computed based on the corresponding parameter estimates and the universal choice set are reported in the last column of Table \ref{tab:book-estimate}, where the value for the SC-Base are too small to be numerically reported, so we indicate it by ``--''. Overall, we can see that the LMDC remarkably outperforms the other baselines, in terms of in-sample fit.

For the out-of-sample evaluation, similar to the Jewelry dataset, we split the set of observations to different subsets according to the values of $[L,U]$. The predicted log-likelihood values are reported in Table \ref{tab:book-prediction} and we indicate in bold the best values across the three models considered. It can be observed that when $L=U$, i.e., the number of selected books is fixed,  the predicted log-likelihood values are quite similar across the three models. As the range of $[L,U]$ expands, the LMDC model begins to significantly outperform the other models. In particular, when $[L,U] = [1,70]$, the average predicted log-likelihood value given by LMDC is -11.51, which is much larger than the value returned by the MC-Base. Moreover, in this case, the predicted log-likelihood of the SC-Base is even too small to be numerically reported and we indicate it by ``--''. In general, similar to the findings observed in the case of the Jewelry dataset, the LMDC model significantly performs better than the other baselines in our out-of-sample evaluation.




\begin{table}
\centering
\caption{Out-of-sample results for the Book-Crossing dataset}
\label{tab:book-prediction}
\begin{tabular}{c|c|crr}
${[}L,U]$ & \begin{tabular}[c]{@{}c@{}}Number of \\observations\end{tabular} & LMDC & \multicolumn{1}{c}{MC-Base} & \multicolumn{1}{c}{SC-Base} \\ 
\hline
$[1,1]$ & 11331 & \textbf{-4.34} & -\textbf{4.34} & \textbf{-4.34} \\
$[2,2]$ & 2275 & \textbf{-8.29} & -8.39 & \textbf{-8.29} \\
$[3,3]$ & 1000 & \textbf{-11.69} & -12.00 & \textbf{-11.69} \\
$[4,4]$ & 575 & \textbf{-14.83} & -15.18 & -14.84 \\
$[5,5]$ & 342 & \textbf{-17.68} & -18.14 & \textbf{-17.68} \\
$[6,6]$ & 256 & \textbf{-20.47} & -20.90 & \textbf{-20.47} \\
$[7,7]$ & 191 & \textbf{-22.99} & -23.50 & -23.00 \\
$[8,8]$ & 144 & \textbf{-25.51} & -25.95 & \textbf{-25.51} \\
$[9,9]$ & 104 & \textbf{-27.83} & -28.27 & \textbf{-27.83} \\
$[2,9]$ & 4887 & \multicolumn{1}{r}{\textbf{-14.10}} & -27.84 & -99.70 \\
$[10,20]$ & 518 & \multicolumn{1}{r}{\textbf{-39.34}} & -48.01 & -126.50 \\
$[1,70]$ & 16986 & \multicolumn{1}{r}{\textbf{-11.51}} & -47.55 & --
\end{tabular}
\end{table}


\subsection{Nested LMDC with Real-World Datasets}
As discussed earlier, the nested RL model can be applied to the DAGs to capture the complex correlation between composite alternatives' utilities. This section presents experimental evaluations of the nested RL model using the two aforementioned real-world datasets. It is worth mentioning that, unlike the RL model, the application of the nested RL to the BiC and MuC DAGs offers two distinct models. Therefore, the results obtained from both BiC and MuC graphs will be presented and compared.  These models are referred to as nested LMDC models.

We employ the Jewelry-Store and Book-Crossing datasets, as in the previous experiments. The attributes used to define the utility functions in the RL model are retained for this experiment as well. The scale functions $\mu_k$, for all $k\in \cN$, are important components of the nested RL model and we specify them as follows. For the Jewelry-Store dataset, for each node $p^c_i$ of the DAG, we define scale function $\mu_{p^c_i}$ as exponential functions of three attributes: \textit{Quantity}, \textit{Const}, and \textit{Count}, where  \textit{Quantity} is the quantity of item $i\in [m]$,  \textit{Const} is a constant attribute taking values of 1, and \textit{Count} is equal to $c$ to capture the number of selected items at that node. The use of exponential functions to model the scales is to guarantee that $\mu_k>0$ for all $k\in \cN$, as suggested by \cite{MaiFosFre15}. Regarding the Book-Crossing data, the scale functions are specified using two attributes: \textit{Const} and \textit{Count}.





\begin{table}[htbt]
    \centering
    \caption{Estimation results of nested LMDC models for the Jewelry-Store dataset}
    \label{tab:jewelry-estimate-nested}
    \resizebox{\textwidth}{!}{
    \begin{tabular}{c|l|rrrr|rrr|c}
    \multirow{2}{*}{DAG}   & \multicolumn{1}{c|}{\multirow{2}{*}{Property}} & \multicolumn{4}{c|}{Utility function}           & \multicolumn{3}{c|}{Scale function }& \multirow{2}{*}{$\mathcal{L}(\widehat{\bbt})$} \\ \cline{3-9}
       & \multicolumn{1}{c|}{}    & \multicolumn{1}{c}{Quantity} & \multicolumn{1}{c}{Price} & \multicolumn{1}{c}{Gender} & \multicolumn{1}{c|}{Const} & \multicolumn{1}{c}{Quantity} & \multicolumn{1}{c}{Const} & \multicolumn{1}{c|}{Count} &       \\ \hline
    \multirow{3}{*}{BiC DAG} & $\widehat{\bbt}$       & 0.785  & 0.066                     & 0.054& -4.062                     & 0.451  & -0.308                    & -0.294    & \multirow{3}{*}{-257,657}                \\
       & Std. Err.     & 0.041  & 0.078                     & 0.015& 0.052& 0.006  & 0.013                     & 0.002     &       \\
       & $t$-test(0)   & 19.088 & 0.842                     & 3.503& -77.677                    & 71.525 & -24.027                   & -129.013  &       \\ \hline
    \multirow{3}{*}{MuC DAG} & $\widehat{\bbt}$       & 1.924 & -0.016 & 0.049 & -3.043 & -0.146   & -1.115  & 0.074    & \multirow{3}{*}{-249,290}                \\
    & Std. Err.     & 0.027  & 0.044 & 0.011 & 0.032 & 0.010   & 0.011  & 0.001    &       \\
    & $t$-test(0)   & 71.153  & -0.365  & 4.546 & -94.506 & -14.600   & -98.057  & 68.516    &       \\ 
    \end{tabular}
    }
\end{table}

\begin{table}[htb]
    \centering
\caption{Parameter estimation of the nested LMDC models for the Book-Crossing dataset.}
    \label{tab:books-estimate-nested}
    \resizebox{\textwidth}{!}{
    \begin{tabular}{c|l|rrrr|rr|c}
    \multirow{2}{*}{Model}   & \multicolumn{1}{c|}{\multirow{2}{*}{Property}} & \multicolumn{4}{c|}{Utility function}           & \multicolumn{2}{c|}{Scale function}                   & \multirow{2}{*}{$\mathcal{L}(\widehat{\bbt})$} \\ \cline{3-8}
       & \multicolumn{1}{c|}{}    & \multicolumn{1}{c}{rating} & \multicolumn{1}{c}{Popularity} & \multicolumn{1}{c}{Age} & \multicolumn{1}{c|}{const} & \multicolumn{1}{c}{Const} & \multicolumn{1}{c|}{Count} &       \\ \hline
    \multirow{3}{*}{BiC DAG} & $\widehat{\bbt}$       & -1.392                     & 2.227    & 0.269                   & -4.505                     & 0.232                     & 0.034& \multirow{3}{*}{-181,020}                \\
       & Std. Err.     & 0.008& 0.008    & 0.008                   & 0.008& 0.002                     & 0.000&       \\
       & $t$-test(0)   & -181.955                   & 289.699  & 35.019                  & -597.270                   & 105.135                   & 1255.165                   &       \\ \hline
    \multirow{3}{*}{MuC DAG} & $\widehat{\bbt}$       & 0.727& 1.670    & -0.121                  & -2.747                     & -0.691                    & 0.034& \multirow{3}{*}{-177,907}                \\
       & Std. Err.     & 1.260& 0.061    & 0.270                   & 1.721& 0.279                     & 0.001&       \\
       & $t$-test(0)   & 0.577& 27.474   & -0.449                  & -1.596                     & -2.477                    & 31.051                     &       \\ 
    \end{tabular}
    }    
\end{table}

The estimation results of the Jewelry-Store and Book-Crossing datasets are reported in Tables \ref{tab:jewelry-estimate-nested} and \ref{tab:books-estimate-nested}. For the Jewelry-Store dataset, similar to the observations we have in the MDLC model, the estimates for \textit{Quantity} and \textit{Count} are significant and have their expected signs, while it is not the case for the estimates of \textit{Price} and \textit{Gender}. The parameter estimates for the three attributes \textit{Quantilty}, \textit{Const} and \textit{Count} are also significant. The final log-likelihood values are presented in the last column of Table \ref{tab:jewelry-estimate-nested}. These values indicate that the nested RL model exhibits superior performance in terms of in-sample fit when applied to the MuC DAG compared to the BiC DAG.

For the Book-Crossing dataset, the results reported in Table \ref{tab:books-estimate-nested} show that all the parameter estimates given by the nested RL model with the BiC DAG are all significant, but the sign of the estimate of \textit{Rating} is negative,  which is intuitively incorrect. On the other hand, the estimates given by the MuC DAG seem to have their expected signs, but they are only significant for \textit{Popularity} of the utility functions, and \textit{Count} of the scale functions. The MuC graph also gives a better in-sample log-likelihood value. In general, we can see that the nested LMDC model based on the MuC DAG offers better performance in terms of in-sample fit. Intuitively, this would be due to the fact that each node in the MuC DAG has several outgoing arcs connecting it with other subsequent nodes while each node in the BiC DAG only has 2 outgoing arcs corresponding to a ``selecting'' or ``'not-selecting'' option,
thus the nested RL on the MuC would be better in capturing the correlation between alternatives.

We now switch ourselves to the out-of-sample evaluation. Table \ref{tab:prediction-nested} reports the average predicted log-likelihood values for both Jewelry-Store and Book-Crossing datasets. For the sake of comparison, we include the average predicted log-likelihood value given by the LMDC model reported previously. The results are computed based on the entire observation sets we have, which range in size from 1 to 70 for the Jewelry-Store dataset and from 1 to 11 for the Book-Crossing dataset. The results clearly indicate that the nested LMDC models, utilizing either the BiC or MuC graphs, outperform the LMDC model for both datasets. Furthermore, the nested LMDC model consistently exhibits better performance when applied to the MuC DAG compared to the BiC DAG.

\begin{table}[htb]
\centering
\caption{Prediction results of the nested LMDC models for the Jewelry-Store and Book-Crossing datasets.}
\label{tab:prediction-nested}
\begin{tabular}{l|l|l|ll}
\multirow{2}{*}{Dataset} & \multirow{2}{*}{${[}L,U]$} & \multirow{2}{*}{LMDC} & \multicolumn{2}{c}{Nested LMDC} \\ 
\cline{4-5}
 &  &  & \multicolumn{1}{c}{BiC DAG} & \multicolumn{1}{c}{MuC DAG} \\ 
\hline
Jewelry-Store & {[}1,11] & -7.24 & -7.14 & -7.15 \\
Book-Crossing & {[}1,70] & -11.51 & -10.37 & -10.28
\end{tabular}
\end{table} 

In summary, our numerical results based on the two real-world datasets indicate several advantages of our modeling and estimation approaches. Firstly,  the LMDC model performs better than the other simple baselines (SC-Base and MC-Base) in terms of both in-sample and out-of-sample fits. In addition, when the nested RL is brought into the context, we also observe that it significantly improves the LMDC model in both in-sample and out-of-sample fits. 

Finally, we report in Table \ref{tab:time-nested} the estimation times of the LMDC and nested LMDC models, for both Jewelry-Store and Book-Crossing datasets. As expected, the computing times of the LMDC model are significantly smaller than those of the nested model. For the LMDC, the estimation times are quite similar across the BiC and MuC graphs. On the other hand, for the nested LMDC models, the MuC DAG requires less computing time, compared to the BiC DAG, even though the MuC is denser. This would be because the nested RL with the  MuC fits better with the data, thus the MLE requires fewer iterations to converge.  In general, for both real datasets with reasonably large numbers of items and observations, our estimation method manages to estimate the  LMDC model in less than 6 minutes and estimate the nested LMDC models in about 1 hour (slightly more or less). This is possible thanks to the use of several techniques and findings employed to estimate the RL and nested RL model mentioned above. In particular, all the computations are formulated as \textit{sparse matrix operations}, which also greatly speeds up the estimation. 

\begin{table}[htb]\small
\centering
\caption{Estimation times (in seconds) for the Jewelry-Store and Book-Crossing datasets .}
\label{tab:time-nested}
\begin{tabular}{l|l|l|c|ll|ll}
\multirow{2}{*}{Dataset} & \multirow{2}{*}{$m$} & \multirow{2}{*}{$[L,U]$} & \multirow{2}{*}{\begin{tabular}[c]{@{}c@{}}No \\observations\end{tabular}} & \multicolumn{2}{c|}{LMDC} & \multicolumn{2}{c}{Nested LMDC} \\ 
\cline{5-8}
 &  &  &  & \multicolumn{1}{c}{BiC DAG} & \multicolumn{1}{c|}{MuC DAG} & \multicolumn{1}{c}{BiC DAG} & \multicolumn{1}{c}{MuC DAG} \\ 
\hline
Jewelry-Store & 436 & {[}1,11] & \multicolumn{1}{l|}{34291} & 333 & 332 & 4134 & 2009 \\
Book-Crossing & 100 & {[}1,70] & \multicolumn{1}{l|}{16986} & 53 & 70 & 1,985 & 926
\end{tabular}
\end{table}


\section{Conclusion}\label{sec:concl}
In this paper, we have conducted a study on models based on RUM for analyzing multiple discrete choices. To address the challenge of having exponentially many alternatives in composite choice sets, we have introduced an innovative approach based on network representations and recursive route choice models. We have proposed two DAGs to capture the multiple discrete choice processes. Furthermore, we have demonstrated that by applying the RL model to these DAGs, we can obtain an equivalent model, implying that the estimation of the LMDC model can be achieved by estimating RL models on the DAGs.

We have also delved into the estimation of the RL model on the DAGs and established important results, such as the polynomial-time computation of the value function using value iteration. Additionally, we have shown that the nested RL approach can be employed to relax the IIA limitation of the LMDC model and capture the correlation between composite alternatives. Theoretical analysis and examples have revealed that applying the nested RL to the BiC and MuC DAGs leads to two distinct models. However, under certain conditions of the scale function, these resulting models can be equivalent.

Our numerical results, based on both synthetic and real datasets, have demonstrated the superior performance of our LMDC model compared to other baselines in terms of in-sample and out-of-sample fits. Remarkably, the estimation of the RL model on the DAGs can be done rapidly, with estimation times within seconds even for instances involving 50 items. We have also provided numerical evaluations using real datasets for the nested LMDC model, which have indicated its improved performance over the LMDC model in both in-sample and out-of-sample fits.

Our research provides innovative and efficient methods to tackle the multi-discrete choice problem through the utilization of route choice models, which opens up several promising avenues for further exploration. For instance, it would be interesting to extend our approach to problems that involve continuous variables or complex behavioral constraints. Additionally, leveraging models and techniques from the route choice literature could prove valuable in addressing challenges related to data availability, such as missing or noisy data. 

\section*{Acknowledgments}
The work is  supported by  the National Research
Foundation Singapore and DSO National Laboratories under the AI Singapore Programme (AISG Award No: AISG2-
RP-2020-017).

\bibliographystyle{plainnat_custom}
\bibliography{refs}

\clearpage
\appendix
\begin{center}
    {\Huge Appendix}
\end{center}

\section{Missing Proofs}

\subsection{Proof of Proposition \ref{prop:size-BiC} }

The complementary counting approach is used to determine the number of nodes. First, we count nodes in a complete BiC DAG without any restriction on the subset size. Each tier $j$ in the first $m+1$ middle tiers has exactly $j+1$ nodes: $\{p_j^0, p_j^1, ..., p_j^j\}$. Add one node $d$ from the final tier, the total number of nodes is
\[
     |\cN_{full}| = \sum_{j=0}^{m} (j+1) + 1 = \frac{m^2}{2} + \frac{3m}{2} + 2
\]


Next, we subtract nodes of form $p_j^c$ that are lost due to violation the constraint $c\leq U$:
\[
    |\cN_U| = \sum_{j=U+1}^m (j-U) = \frac{m^2}{2} + \frac{m}{2} - mU + \frac{U^2}{2} - \frac{U}{2}
\]

Aggregation of these two components leads to the final node count in the BiC DAG:
\begin{align*}
\begin{split}
    |\cN|   
    & = |\cN_{full}| - |\cN_U| \\
    & = m + mU - \frac{U^2}{2} + \frac{U}{2} + 2.
\end{split}
\end{align*}

For the number of arcs, we see that most nodes have 2 outgoing arcs, except $U$ nodes belonging to the $m$-th tier and the absorbing node $d$. The first $L-1$ nodes of the $m$-th tier have no arc because they don't satisfy minimum requirement $L$. The remaining $U - L + 1$ nodes have exactly 1 arc leading to $d$. Another exception is the absorbing state $d$ which has no arc. As a result, the number of arcs in a BiC DAG is
\begin{align}\nonumber
\begin{split}
    |\cA| 
    &= 2 (|\cN| - U - 1) + (U - L + 1) \\
    &= 2m + 2mU - U^2 - L + 1. 
\end{split}
\end{align}

\subsection{Proof of Proposition \ref{prop:number-node-MuC}}

The node structure in MuC DAG is similar to BiC DAG's, except all $m$ nodes of form $p_j^0$ are lost:
\begin{align}\nonumber
\begin{split}
    |\cN|   
   = mU - \frac{U^2}{2} + \frac{U}{2} + 2
\end{split}
\end{align}

For the number of arcs, we first consider $L=0$ as a special case when counting the number of arcs in the graph. Another interesting observation is that the total number of arcs in the cases of $L=0$ and $L=1$ differs by exactly one edge connecting the source node to the destination node. Therefore, in the subsequent computations, we temporarily consider $L\geq 1$.

We denote the superscript $c$ of $p_j^c$ as their ``depths''. Due to the size constraint, the depth of a node cannot exceed $U$. Arcs in the  MuC DAG can be divided into three categories based on depths: arcs start from the source node $p^0_0$, outgoing arcs from nodes at depths $c < L$, and arcs from the remaining nodes (from depth $L$ to $U$). The edge count in each category is determined as follows:
\begin{itemize}
    \item {
    Arcs from the node $p^0_0$: In case $L\geq 1$, there are only $|\cA_1| = m$ arcs from $p^0_0$ to nodes at depth $c=1$:
    }
    \item {
    Arcs from nodes at depths $c < L$: Each depth $c$ contains $m-c+1$ nodes. Each node $p_j^c$ at depth $c$ has arcs leading to some nodes of form $p_{j'}^{c+1}$ where $j<j'\leq m$. In conclusion, the edge count in this scenario is
    \[
         |\cA_2| = \sum_{c=1}^{L-1} \sum_{j=c}^{m} \sum_{j'=j+1}^{m} 1
    \]
    }
    \item {
    Arcs from nodes at depths $c\in[L, U]$: Each depth $c$ contains $m-c+1$ nodes, and each node $p_j^c$ has arcs leading to nodes of the form $p_{j'}^{c+1}$ where $j<j'\leq m$ plus one arc to the final node $d$. The total arc count is
    \[
         |\cA_3| = \sum_{c=L}^{U} \sum_{j=c}^{m} (\sum_{j'=j+1}^{m} 1 + 1)
    \]
    }
\end{itemize}
So, the total number of arcs in the MuC DAG is:
\begin{align}\nonumber
\begin{split}
    |\cA|   
    &= |\cA_1| + |\cA_2| + |\cA_3| \\
    &= \frac{m^2U}{2} - \frac{mU^2}{2} - mL + mU + 2m + \frac{L^2}{2} - \frac{3L}{2} + \frac{U^3}{6} - \frac{U^2}{2} + \frac{U}{3} + 1.
\end{split}
\end{align}
We complete the proof.

\subsection{Proof of Proposition \ref{prop:nested-util}}
We first prove the following lemma showing that the value functions for both BiC and MuC are identical.
\begin{lemma}
    Under the assumptions of Proposition \ref{prop:number-node-MuC}, the value functions are the same for both graphs, i.e.,  for two nodes $ k\in \cN^\BiC, k'\in \MuC$ such that $k$ and $k'$ are both of the same form $p^c_i$, for $c\in [0,U], i\in 0,\ldots,m$, then 
\[
\widetilde{V}^{\BiC}(k) = \widetilde{V}^{\MuC}(k').\]
\end{lemma}
\proof
For notational simplification, we write $\widetilde{V}^\BiC(p^c_j)$ to denote the value function at the node of form $p^c_i$ in the BiC graph, and $\widetilde{V}^\MuC(p^c_j)$ denotes the value function at the node of form $p^c_j$ in the MuC graph. Moreover, from the assumption that the scale function at node $p^c_j$ is only dependent on $c$, to facilitate the later exposition, let us define $\mu^*(c)$, for any $c\in [0,U]$ such that $\mu_{p^c_j} = \mu^*(c)$.  
Moreover,  to simplify the proofing process, we assume there is no bounding constraint related to the number of selected alternatives. This means there is no arc removed due to constraint conflicts in both DAGs. Given the stated conditions, we will prove the equality between $\widetilde{V}^{\BiC}(p_j^k)$ and $\widetilde{V}^{\MuC}(p_j^k)$ by induction through all $m+2$ tiers of nodes in reversed order (from Tier $m+1$ to Tier 0) in both BiC DAG and MuC DAG. 

First, the result holds for the absorbing state $d$ of the final tier since $\widetilde{V}^{\BiC}(d) = \widetilde{V}^{\MuC}(d) = 0$. Afterwards, we assume $\widetilde{V}^{\BiC}(p_{j'}^k) = \widetilde{V}^{\MuC}(p_{j'}^k)$ for all nodes lying from tier $j+1$ to the final tier $m+1$. Our goal is to prove this statement for tier $i$: $\widetilde{V}^{\BiC}(p_j^k) = \widetilde{V}^{\MuC}(p_j^k)$.

In the nested BiC DAG, each node $p_j^k$ leads to two following nodes $p_{j+1}^k$ and $p_{j+1}^{k+1}$ (except nodes only leading to $d$). Based on \autoref{eq:nested-value}, the expected maximum utility of each node $p_j^k$ can be recursively derived from the system

\begin{equation*}
    \exp{\left( \frac{1}{\mu_{p_j^k}} \widetilde{V}^\BiC(p_j^k) \right)} = 
    \begin{cases}
    1,   & j=m,\, k\in [L,U] \\
    0,   & j=m,\, k\notin [L,U] \\
    \!\begin{multlined}[t]
    \exp{\left( \frac{1}{\mu_{p_j^k}} (v^\BiC(p_{j+1}^k|p_j^k) + \widetilde{V}^\BiC(p_{j+1}^k)) \right)} \\
    + \exp{\left( \frac{1}{\mu_{p_j^k}} (v^\BiC(p_{j+1}^{k+1}|p_j^k) + \widetilde{V}^\BiC(p_{j+1}^{k+1})) \right)}
    \end{multlined}
       & 0\leq j<m
    \end{cases}
\end{equation*}

Since the arc $(p_j^k,p_{j+1}^k)$ represents the ``not selecting'' option, $v^\BiC(p_{j+1}^k|p_j^k)=0$. The other arc provides the equation $v^\BiC(p_{j+1}^k|p_j^k) = v_{j+1}$. On the other hand, $\mu_{p_j^k}$ can be replaced with $\mu^*(k)$ from the given statements. For $p_j^k$ nodes where $j<m$, the equation becomes
\begin{equation*}
    \exp{\left( \frac{1}{\mu^*(k)} \widetilde{V}^\BiC(p_j^k) \right)} =
    \!\begin{multlined}[t]
    \exp{\left( \frac{1}{\mu^*(k)} (v_{j+1} + \widetilde{V}^\BiC(p_{j+1}^{k+1})) \right)} \\
    + \exp{\left( \frac{1}{\mu^*(k)} \widetilde{V}^\BiC(p_{j+1}^k) \right)}
    \end{multlined}
\end{equation*}

Apply the above formulation on node $p_{j+1}^k$, we get
\begin{equation*}
    \exp{\left( \frac{1}{\mu^*(k)} \widetilde{V}^\BiC(p_{j+1}^k) \right)} =
    \!\begin{multlined}[t]
    \exp{\left( \frac{1}{\mu^*(k)} (v_{j+2} + \widetilde{V}^\BiC(p_{j+2}^{k+1})) \right)} \\
    + \exp{\left( \frac{1}{\mu^*(k)} \widetilde{V}^\BiC(p_{j+2}^k) \right)}
    \end{multlined}
\end{equation*}

which leads to the following if combined with the previous equation
\begin{equation*}
    \exp{\left( \frac{1}{\mu^*(k)} \widetilde{V}^\BiC(p_j^k) \right)} =
    \!\begin{multlined}[t]
    \exp{\left( \frac{1}{\mu^*(k)} (v_{j+1} + \widetilde{V}^\BiC(p_{j+1}^{k+1})) \right)} \\
    + \exp{\left( \frac{1}{\mu^*(k)} (v_{j+2} + \widetilde{V}^\BiC(p_{j+2}^{k+1})) \right)} \\
    + \exp{\left( \frac{1}{\mu^*(k)} \widetilde{V}^\BiC(p_{j+2}^k) \right)}
    \end{multlined}
\end{equation*}

This replacement action can be recursively applied until a determinable state is reached
\begin{equation*}
    \exp{\left( \frac{1}{\mu^*(k)} \widetilde{V}^\BiC(p_m^k) \right)} = 
    \begin{cases}
    1,   & k\in [L,U] \\
    0,   & k\notin [L,U]
    \end{cases}
\end{equation*}

Therefore, the final equation achieved is
\begin{equation*}
    \exp{\left( \frac{1}{\mu^*(k)} \widetilde{V}^\BiC(p_j^k) \right)} = 
    \begin{cases}
    \sum_{j'=j+1}^m \exp{\left( \frac{1}{\mu^*(k)} (v_{j'} + \widetilde{V}^\BiC(p_{j'}^{k+1})) \right)} + 1,   & k\in [L,U] \\
    \sum_{j'=j+1}^m \exp{\left( \frac{1}{\mu^*(k)} (v_{j'} + \widetilde{V}^\BiC(p_{j'}^{k+1})) \right)},   & k\notin [L,U]
    \end{cases}
\end{equation*}

Fortunately, according to the DAG topology in \autoref{fig:choice-graph}, expected utility values in the MuC DAG follow a similar system

\begin{equation*}
    \exp{\left( \frac{1}{\mu_{p_j^k}} \widetilde{V}^\MuC(p_j^k) \right)} = 
    \begin{cases}
    1,   & j=m,\, k\in [L,U] \\
    0,   & j=m,\, k\notin [L,U] \\
    \sum_{j'=j+1}^m \exp{\left( \frac{1}{\mu^*(k)} (v_{j'} + \widetilde{V}^\MuC(p_{j'}^{k+1})) \right)} + 1,   & 0\leq j<m,\, k\in [L,U] \\
    \sum_{j'=j+1}^m \exp{\left( \frac{1}{\mu^*(k)} (v_{j'} + \widetilde{V}^\MuC(p_{j'}^{k+1})) \right)},   & 0\leq j<m,\, k\notin [L,U]
    \end{cases}
\end{equation*}

The only exceptions are nodes of form $p_j^0,~\forall j|0<j\leq m$ which don't exist in MuC DAG. Nevertheless, when $\widetilde{V}^\BiC(p_{j'}^k)=\widetilde{V}^\MuC(p_{j'}^k),~ \forall j'|j<j'\leq m$, it is easy to also point out the equality of nodes in tier $i$ also holds: $\widetilde{V}^\BiC(p_j^k)=\widetilde{V}^\MuC(p_j^k)$. As such, the hypothesis is successfully proved by induction.

\endproof

We are now to prove that the probability distribution over paths given by both BiC and MuC are also the same. 
To this end, we see that the probability of each path is the product of all arcs' probabilities. In both nested BiC and MuC models, the probability of an arc from state $k$ to state $a$ can be rewritten as:
\[
    \widetilde{P}(a|k) = \frac{\exp{\left(\frac{1}{\mu_k} v(a|k)\right)} \exp{\left(\frac{1}{\mu_k} \widetilde{V}(a)\right)}}{\exp{\left(\frac{1}{\mu_k} \widetilde{V}(k)\right)}}.
\]
Next, we investigate the probability of a path $\sigma^\BiC(S)$ corresponding to  subset $S = \{j_1,j_2,\ldots,j_K\}$ where $j_1<j_2<\ldots<j_K$. Proposition~\ref{prop:map-BiC} denotes $\sigma^{\BiC}(S)$ as a path consisting of $K+1$ segments. The $k^{th}$ segment runs from $p_{j_k}^k$ to $p_{j_{k+1}-1}^k$: $\sigma^\BiC_k(S) = \{p_{j_k}^k, p_{j_k+1}^k,\ldots, p_{j_{k+1}-1}^k\}$ such that the $0^{th}$ one has $j_0=0$. The last $K^{th}$ segment starts at $p_{j_K}^K$ and ends at $d$: $\sigma^\BiC_K(S) = \{p_{j_K}^K, p_{j_K+1}^K,\ldots, p_m^K, d\}$. The probability of the $k^{th}$ segment ($0\leq k<K$) is
\[
    \widetilde{P}^{\BiC}(\sigma^{\BiC}_k(S)) = \prod_{j=j_k}^{j_{k+1}-2} \frac{\exp{\left(\frac{1}{\mu_{p_j^k}} v(p_{j+1}^k|p_j^k)\right)} \exp{\left(\frac{1}{\mu_{p_j^k}} \widetilde{V}^{\BiC}(p_{j+1}^k)\right)}}{\exp{\left(\frac{1}{\mu_{p_j^k}} \widetilde{V}^{\BiC}(p_j^k)\right)}}
\]
Since all arcs in a segment corresponds to an action of not selecting an alternative (and doesn't increase the selection count $k$), all utility values are zero: $\exp{\left(\frac{1}{\mu_k} v(p_{j+1}^k|p_j^k)\right)} = 1$. The considered scenario also provide a formula for the scale parameters: $\mu_{p_j^k} = \mu^*(k)$. Replace these alternative forms and reduce, we have
\[
    \widetilde{P}^{\BiC}(\sigma^{\BiC}_k(S)) = \prod_{j=j_k}^{j_{k+1}-2} \frac{\exp{\left(\frac{1}{\mu^*(k)} \widetilde{V}^{\BiC}(p_{j+1}^k)\right)}}{\exp{\left(\frac{1}{\mu^*(k)} \widetilde{V}^{\BiC}(p_j^k)\right)}} = \frac{\exp{\left(\frac{1}{\mu^*(k)} \widetilde{V}^{\BiC}(p_{j_{k+1}-1}^k)\right)}}{\exp{\left(\frac{1}{\mu^*(k)} \widetilde{V}^{\BiC}(p_{j_k}^k)\right)}}
\]

In the case of the $K^{th}$ segment, the last arc from $p_m^K$ to $d$ also has zero utility. Moreover, $\widetilde{V}^{\BiC}(d) = 0$. Therefore:
\[
    \widetilde{P}^{\BiC}(\sigma^{\BiC}_K(S)) = \left( \prod_{j=j_K}^{m-1} \frac{\exp{\left(\frac{1}{\mu^*(k)} \widetilde{V}^{\BiC}(p_{j+1}^K)\right)}}{\exp{\left(\frac{1}{\mu^*(K)} \widetilde{V}^{\BiC} (p_j^K)\right)}} \right) \frac{1}{\exp{\left(\frac{1}{\mu^*(K)} \widetilde{V}^{\BiC} (p_m^K)\right)}}
    = \frac{1}{\exp{\left(\frac{1}{\mu^*(K)} \widetilde{V}^{\BiC} (p_{j_K}^K)\right)}}
\]

We multiply the probability of all segments to achieve the total likelihood of the complete path. Note that between two consecutive segments $k$ and $k+1$ there exists an intermediate arc $(p_{j_{k+1}-1}^k, p_{j_{k+1}}^{k+1})$ with utility $v_{k+1}$ and scale parameters $\mu_{p_{j_{k+1}-1}^k} = \mu^*(k)$.

\begin{equation}\label{eq:bic-nested-path}
\begin{aligned}
    \widetilde{P}^{\BiC}(\sigma^{\BiC}(S))
    &= \prod_{k=0}^{K-1} \left(\widetilde{P}^{\BiC}(\sigma^{\BiC}_k(S))  
    \frac{\exp{\left(\frac{1}{\mu^*(k)} v_{k+1}\right)} \exp{\left(\frac{1}{\mu^*(k)} \widetilde{V}^{\BiC}(p_{j_{k+1}}^{k+1})\right)}}
    {\exp{\left(\frac{1}{\mu^*(k)} \widetilde{V}^{\BiC}(p_{j_{k+1}-1}^k)\right)}} \right) 
    \frac{1}{\exp{\left(\frac{1}{\mu^*(K)} \widetilde{V}^{\BiC} (p_{j_K}^K)\right)}} \\
    &= \prod_{k=0}^{K-1} \frac{\exp{\left(\frac{1}{\mu^*(k)} v_{k+1}\right)} \exp{\left(\frac{1}{\mu^*(k)} \widetilde{V}^{\BiC}(p_{j_{k+1}}^{k+1})\right)}}
    {\exp{\left(\frac{1}{\mu^*(k)} \widetilde{V}^{\BiC}(p_{j_k}^k)\right)}} 
    \frac{1}{\exp{\left(\frac{1}{\mu^*(K)} \widetilde{V}^{\BiC} (p_{j_K}^K)\right)}}
\end{aligned}
\end{equation}

Finally, we evaluate the corresponding path $\sigma^{\MuC}(S) = (p^0_0, p^1_{j_1}, p^2_{j_2}, \ldots, p^K_{j_K}, d)$ for the subset $S = \{j_1,j_2,\ldots,j_K\}$ in MuC DAG. Its probability is also the product of all arcs:

\begin{equation*}
\begin{aligned}
    \widetilde{P}^{\MuC}(\sigma^{\MuC}(S))
    &= \left( \prod_{k=0}^{K-1} \widetilde{P}^{\MuC}(p_{j_{k+1}}^{k+1}|p_{j_k}^k) \right)
    \widetilde{P}^{\MuC}(d|p_{j_K}^K) \\
    &= \prod_{k=0}^{K-1} \frac{\exp{\left(\frac{1}{\mu_{p_{j_k}^k}} v(p_{j_{k+1}}^{k+1}|p_{j_k}^k)\right)} \exp{\left(\frac{1}{\mu_{p_{j_k}^k}} \widetilde{V}^{\MuC}(p_{j_{k+1}}^{k+1})\right)}}
    {\exp{\left(\frac{1}{\mu_{p_{j_k}^k}} \widetilde{V}^{\MuC}(p_{j_k}^k)\right)}} 
    \frac{\exp{\left(\frac{1}{\mu_{p_{j_K}^K}} (v(d|p_{j_K}^K) + \widetilde{V}^{\MuC} (d))\right)}}
    {\exp{\left(\frac{1}{\mu_{p_{j_K}^K}} \widetilde{V}^{\MuC} (p_{j_K}^K)\right)}} 
\end{aligned}
\end{equation*}

Each arc has an utility $v(p_{j_{k+1}}^{k+1}|p_{j_k}^k) = v_{k+1}$, except $v(d|p_{j_K}^K) = \widetilde{V}^{\MuC}(d) = 0$. The condition statements give $\mu_{p_{j_k}^k} = \mu^*(k)$. Consequently, the above equation becomes

\begin{equation}\label{eq:muc-nested-path}
    \widetilde{P}^{\MuC}(\sigma^{\BiC}(S))
    = \prod_{k=0}^{K-1} \frac{\exp{\left(\frac{1}{\mu^*(k)} v_{k+1}\right)} \exp{\left(\frac{1}{\mu^*(k)} \widetilde{V}^{\MuC}(p_{j_{k+1}}^{k+1})\right)}}
    {\exp{\left(\frac{1}{\mu^*(k)} \widetilde{V}^{\MuC}(p_{j_k}^k)\right)}} 
    \frac{1}{\exp{\left(\frac{1}{\mu^*(K)} \widetilde{V}^{\MuC} (p_{j_K}^K)\right)}}.
\end{equation}
From Proposition~\ref{prop:nested-util}, we have $\widetilde{V}^{\BiC}(p_j^k) = \widetilde{V}^{\MuC}(p_j^k)$ for all $p_j^c$ nodes that exist in both DAGs. Apply this equality to \autoref{eq:bic-nested-path} and \autoref{eq:muc-nested-path}, we have
\[
    \widetilde{P}^{\BiC}(\sigma^{\BiC}(S)) =   \widetilde{P}^{\MuC}(\sigma^{\MuC}(S)),
\]
as desired.

\section{Extension to Multiple Discrete-Count Choice Problem}
In the main body of the paper, our assumption was that each individual had only two choices: either selecting or not selecting an elemental alternative. In this section, we will explore a scenario where individuals can choose each elemental alternative multiple times. In this particular setup, every observation consists of a set of pairs $(i, c_i)$, where $i\in[m]$ is an elemental alternative,  and $c_i$ is a non-negative integer representing the number of times the alternative $i$ is chosen. The total number of selected alternatives must also fall within the range of $[L, U]$, i.e., $\sum_{i\in [m]}c_i \in [L,U]$. In this section, we will demonstrate how the BiC and MuC DAGs can be adjusted to accommodate this extended scenario.

We first describe the LMDC model adapted for the multiple discrete-count setting. Each composite alternative $S$ now is a set of pairs of elemental alternatives and their counts, i.e., $S = \{(i,c_i)|~ i\in [m]\}$. Let us denote by $\overline{\Omega}^{[L,U]}$  the set of all possible composite alternatives, $$\overline{\Omega}^{[L,U]} = \left\{S = \{(i,c_i)|~ i\in [m]\}\Big| ~ \sum_{i\in[m]} c_i \in [L,U]\right\}.$$ Under the logit setting,
the choice probability of an alternative $S\in\overline{\Omega}^{[L,U]}$ becomes 
\begin{equation}\label{eq:multi-count-eq1}   
P(S|~\overline{\Omega}^{[L,U]}) = \frac{\exp(v(S|\bbt))}{\sum_{S'\in \overline{\Omega}^{[L,U]}}\exp(v(S'|\bbt))},
\end{equation}
where $V(S|\bbt)$ is the total utility of the composite alternative $S$, computed as $V(S|\bbt) = \sum_{(i,c_i)\in S} c_i v_i(\bbt)$. It can be seen that the choice probability formulation given in \eqref{eq:multi-count-eq1} is a generalization of \eqref{eq:multi-choice-RUM} when $c_i$ only takes values of 0 or 1. The log-likelihood function can be defined in the same way as in Section \ref{sec:LMDC}. It can also be proved that the log-likelihood function, in the extended context, is also concave in $\bbt$, guaranteeing global optimality when solving the maximum likelihood estimation. We will describe, in the following, BiC and MuC DAGs for this choice model. 

\textbf{Adapted BiC DAG. } We also create a DAG whose nodes are denoted as $p^c_i$ where the superscript $c$ represents the accumulated number of selected elemental alternatives, and $i\in [m]$ is the index of the corresponding alternative. Nodes are also arranged in $(m+2)$ tiers where each tier will consist of $(U+1)$ nodes representing 
the fact that each elemental alternative can be selected at most $U$ times, from 0 to $U$. In each tier, we create arcs connecting nodes $p^c_i$ with $p^{c+1}_i$, which represents an additional selection of alternative $i$. Across different tiers, we create arcs connecting nodes $p^c_i$ with $p^c_{i+1}$, representing the completion of selecting alternative $i$ and the transition to the next alternative $i+1$. So, from each node, there are two outgoing arcs representing an additional selection of the same alternative or a transition to the next alternative in the sequence. We illustrate the adapted BiC DAG in \autoref{fig:multi-count-BiC} below. 

To apply the RL to the adapted BiC DAG, we define  arc utilities as follows
 \begin{equation}\label{eq:v-adaptedBiC}
\begin{cases}
(a):~ v^{\BiC}(p^c_{i+1}|p^c_i) = 0,~\forall c\in \{0,\ldots,U\},~i \in \{0,...,m\}    \\
(b):~ v^{\BiC}(p^{c+1}_{i}|p^{c}_{i}) = v_i, ~\forall c\in \{0,\ldots,U-1\},~i \in \{0,...,m\}    \\
(c):~v^{\BiC}(k|d) = 0, ~\forall k\in \cN, d\in N(k), 
\end{cases}
\end{equation}
where the assignment $(a)$ is for a transition to the next alternative in the sequence, which does not generate any utilities, and $(b)$ is for a selection of alternative $i$, generating a utility of $v_i$. 

\begin{figure}[htb]
    \centering    \includegraphics[width=0.8\linewidth]{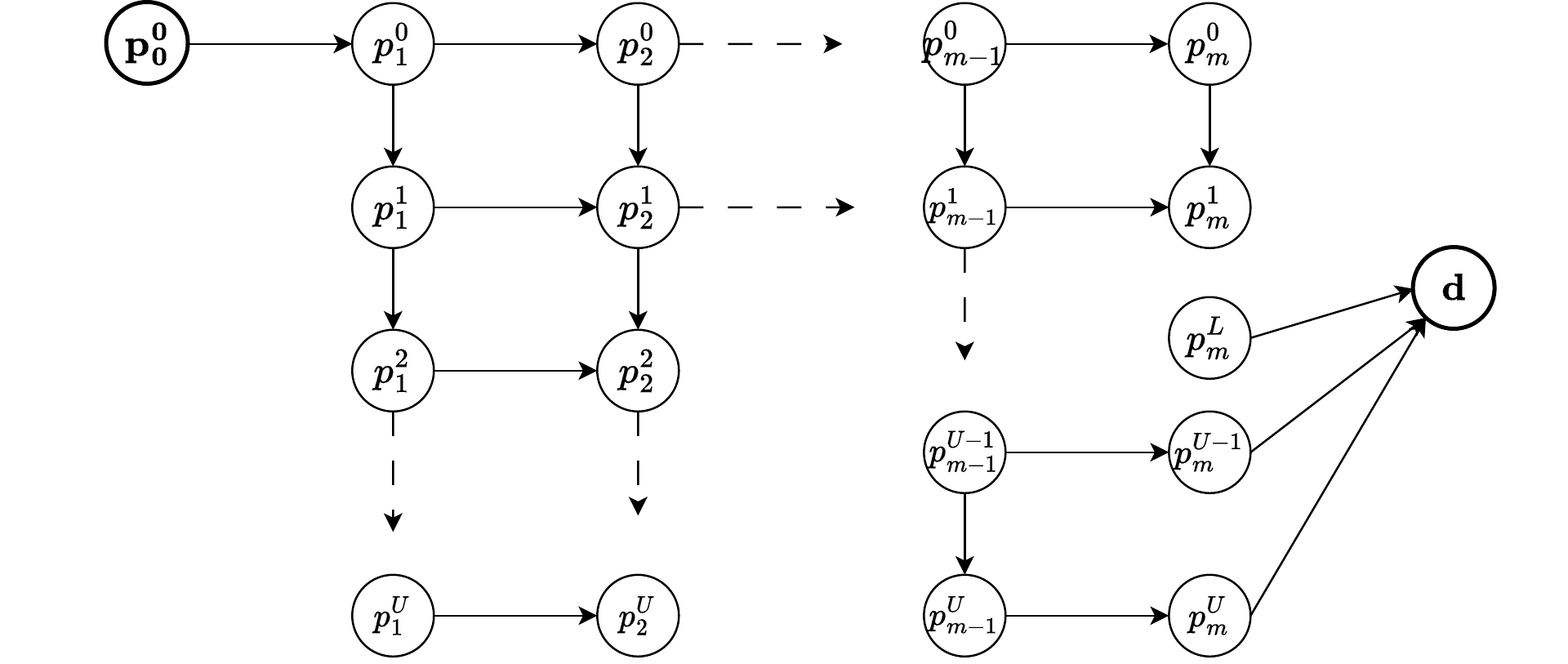}
    \caption{Adapted BiC DAG for multiple discrete-count choices.}
    \label{fig:multi-count-BiC}
\end{figure}

\begin{figure}[htb]
    \centering    \includegraphics[width=0.8\linewidth]{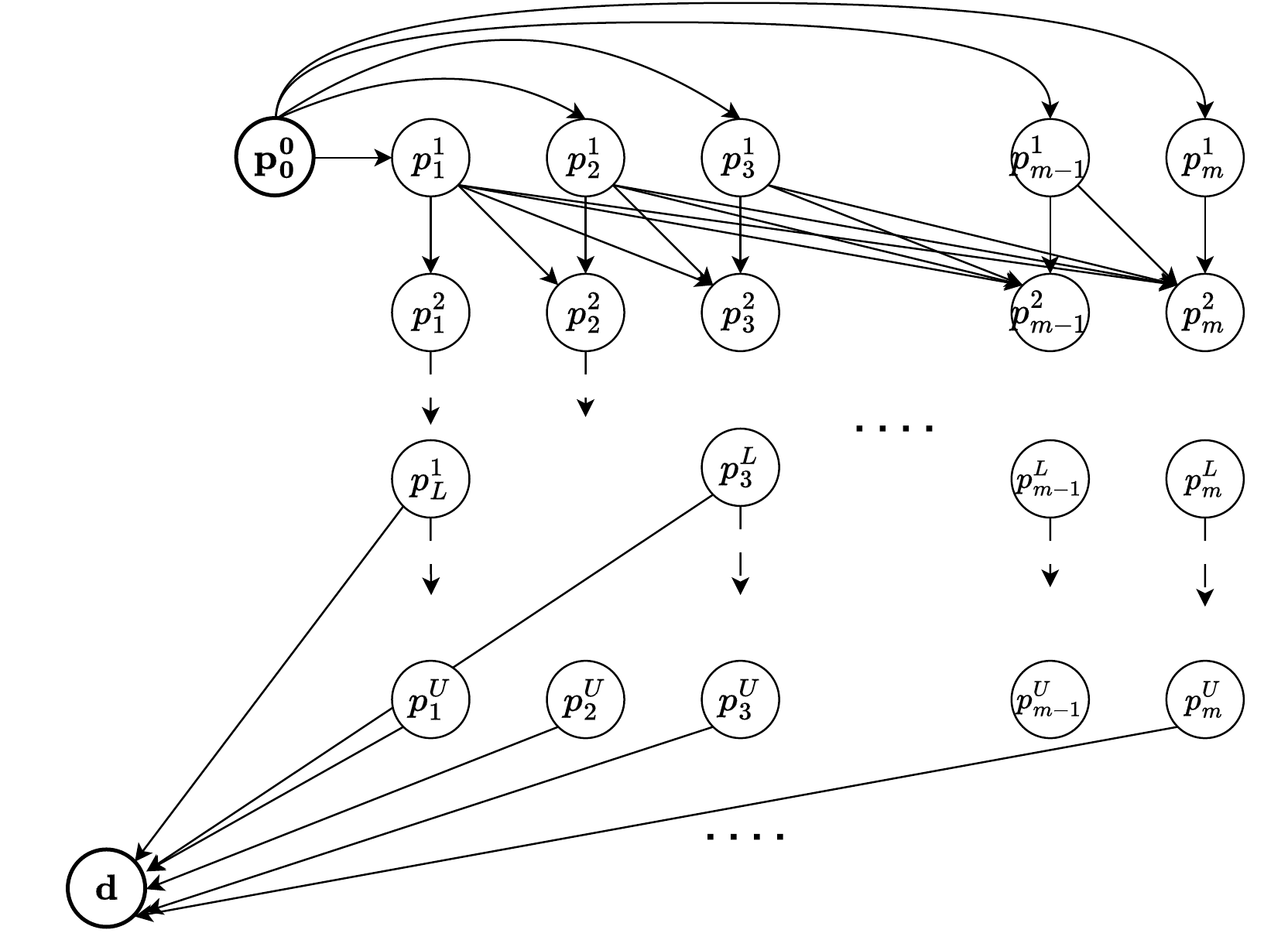}
    \caption{Adapted MuC DAG for multiple discrete-count choices.}
    \label{fig:multi-count-MuC}
\end{figure}

\textbf{Adapted MuC DAG. } We do the same for the MuC DAG. That is, nodes are also organized into $(m+2)$ tiers where the first tier consists of only one node $p^0_0$ representing the start of the choice process, and the last tier consists of one node $d$ representing the termination of the choice process. In Tier $i\in [m]$, we include $(U+1)$ nodes of the form $p^c_i$, where $c = 1,\ldots,U$, representing the fact that each elemental alternative can be selected at most $U$ times.  Within each tier $i\in [m]$, we create arcs between nodes $p^c_i$ and $p^{c+1}_i$, representing an additional selection of alternative $i$. Across tiers, we connect nodes $p^c_i$ with $p^{c+1}_{i+1},p^{c+1}_{i+2},...,p^{c+1}_{m}$, representing selections of next alternatives in the sequence, noting that we only connect nodes $p^c_{i}$, for $c\geq L, i\in [m]$, with the destination $d$. 
The adapted MuC DAG is illustrated in \autoref{fig:multi-count-MuC}. 

For the use of the RL model on the adapted MuC DAG, we define arc utilities as follows:
 \begin{equation}\label{eq:v-adaptedMuC}
\begin{cases}
(a'):~ v^{\MuC}(p^c_{i}|p^{c+1}_i) = v_i,~\forall c\in \{0,\ldots,U-1\},~i \in \{1,...,m\}    \\
(b'):~ v^{\MuC}(p^{c+1}_{i+1}|p^{c}_{i}) = v_{i+1}, ~\forall c\in \{0,\ldots,U-1\},~i \in \{0,...,m-1\}, \\
(c'):~v^{\MuC}(k|d) = 0, ~\forall k\in \cN, d\in N(k), 
\end{cases}
\end{equation}
where $(a')$ are the assignments for arcs that represent a repeated selection of alternative $i\in [m]$, and $(b')$ are for arcs that represent transitions to subsequent alternatives.

Similar to the multiple discrete choice problem considered in the main body of the paper, it can be proved that there is a one-to-one mapping between each discrete-count composite alternative in $\overline{\Omega}^{[L,U]}$
 and a path in the adapted BiC (or adapted MuC) DAG, and when applying the RL model to the adapted graphs,  the path choice distributions over paths in the adapted BiC (or adapted MuC) are the same as the distribution over composite alternatives given by \eqref{eq:multi-count-eq1}. We state these results in the following proposition: 
\begin{proposition}
 The following holds:
 \begin{itemize}
     \item[(i)] There is a one-to-one mapping between each composite alternative in $\overline{\Omega}^{[L,U]}$ and a path in the adapted BiC DAG, and a path in the adapted MuC DAG.
     \item[(ii)] For each $S\in \overline{\Omega}^{[L,U]}$, let $\overline{\sigma}^\BiC(S)$ and $\overline{\sigma}^\MuC(S)$ be its mapped paths in the adapted BiC and MuC DAGs, respectively. In addition, let $\overline{P}^\BiC(\sigma)$ and $\overline{P}^\MuC(\sigma)$ be the probabilities of a given path $\sigma$ given by the RL model applied to the adapted BiC and MuC graphs, respectively, we have
     \[
     P(S|~\overline{\Omega}^{[L,U]}) = \overline{P}^\BiC(\overline{\sigma}^\BiC(S)) = \overline{P}^\MuC(\overline{\sigma}^\MuC(S)).
     \]
     As a result, the multiple discrete-count choice model can be estimated via estimating the RL model on the DAGs.
     \item[(iii)] When computing the value function required during the estimation of the RL or nested RL model, for both adapted BiC and MuC, the value iteration always returns a fixed-point solution after $(m+2)$ iterations.   
 \end{itemize}
\end{proposition}
 The proof is very similar to the proofs of similar results presented in the previous sections. We, therefore, omit it.

\end{document}